%% file: delphi98-174phys813.tex
\documentstyle[12pt,epsf,epsfig,amssymb]{article}
\include{commands}

\begin{document}
%
%
\begin{titlepage}
\pagenumbering{arabic}
\vspace*{-1.5cm}
\begin{tabular*}{15.cm}{l@{\extracolsep{\fill}}r}
{\bf DELPHI Collaboration} & DELPHI 98-174 PHYS 813  \\
                           & WU B 98-39             \\
                           & 7 December, 1998    \\                           
\hline
\end{tabular*}
\vspace*{2.0cm}

\begin{center}
\Large {\bf The Strong Coupling: \\ Measurements and Running}\\
\vspace*{2.0cm}

\normalsize { 
   {\bf S. Hahn}  \\ 
   {\footnotesize Fachbereich Physik, Bergische Universit{\"a}t-GH Wuppertal}\\
   {\footnotesize Gau\ss{}stra\ss{}e 20, 42097 Wuppertal, Germany}\\   
   {\footnotesize e-mail: s.hahn@cern.ch}\\   }
\vspace*{2cm}
\end{center}

\vspace{\fill}

\begin{abstract}
\noindent
Measurements of \as from event shapes in \ee annihilation are 
discussed including recent determinations using experimentally 
optimized scales, studies of theoretically motivated scale
setting prescriptions, and recently observed problems with
predictions in Next to Leading Logarithmic Approximation. 
Other recent precision measurements of \as are briefly 
discussed. The relevance of power terms for the energy 
evolution of event shape means and distributions is 
demonstrated. Finally a summary on the current results 
on \asmz and its running is given.
\end{abstract}

\vspace{\fill}

\begin{center}
Plenary talk presented at the Hadron Structure'98  \\
Stara Lesna, September 7-13, 1998
\end{center}

\vspace{\fill}

\end{titlepage}

\pagebreak


\section{Introduction}

Most of our current knowledge about the strong coupling and its running
comes from the analysis of the process \ee $\rightarrow hadrons$ which
is the simplest possible initial state for strong interaction. The energy
scale of the process is exactly known and hadronic interaction is limited 
to the final state. For many observables the hadronization corrections
are proportional $1/Q$ and therefore LEP with an energy range up to
about $200 \gev $ in the year 2000 is an ideal laboratory for studying QCD.   
Due to the huge cross section at the \ZZ resonance every LEP 
experiment collected several million hadronic events, serving for a 
precise determination of \asmzx. With the high energy data from LEP 2,
the running of \as and the influence of non-perturbative power terms 
on the observed quantities can be studied in detail.  

\hspace{0.2 cm}
 

\section{Consistent Measurements of \bfas from the Analysis of \ZZbf 
         Data using Oriented Event Shapes}

Within the recent analysis \cite{delpap} of \ZZ data by the DELPHI 
collaboration, the distributions of 18 different shape observables are 
determined as a function of the polar angle \tht of the thrust axis 
with respect to the \ee beam direction. 
Since the definition of the thrust axis has a forward-backward 
ambiguity, \cttgz has been chosen, \ctt is called the event orientation.
The definition of the observables studied can be found in \cite{delpap}. 
The theoretical predictions in \oass are calculated using 
EVENT2 \cite{event}, which applies the matrix elements of the 
Leiden Group \cite{leiden}. Using this program,  
the double differential cross section for any IR- and collinear 
safe observable $ Y $ in \ee annihilation in dependence on the event 
orientation can be calculated:

\begin{eqnarray} 
\label{Xsect}
    \frac { 1 } { \sigma_{tot} } 
    \frac { d^{2} \sigma (Y, \cos \vartheta ) } 
          { dY d \cos \vartheta }                 
    &  =                                                   
    &  \bar{\alpha}_s ( \mu^2 ) \cdot A (Y, \cos \vartheta )     \\    
    &  + 
    &  \bar{ \alpha}_s^2 ( \mu^2 ) \cdot       
       \Big[ B (Y, \cos \vartheta ) 
             + ( 2 \pi \beta_{0} \ln ( x_{\mu} ) -2 ) 
             A(Y,\cos\vartheta) \Big]                         \nonumber             
\end{eqnarray} 

\nin 
where $\bar{ \alpha}_s = \alpha_s / 2\pi $ 
and $\rm \beta_0 = (33 - 2n_f) / 12 \pi $. 
$ \sigma_{tot} $ is the one loop corrected cross section for the 
process \ee $\rm \to $ hadrons. $\rm A $ and $\rm B $ denote 
the \oas and \oass QCD coefficient functions, respectively. \\

The dependence on the renormalization scale $ \mu $ enters 
logarithmically in \oassx. The scale factor $x_{\mu} $ is defined by 
$\mu^2 = x_{\mu} Q^2 $ where $Q = M_Z $ is the center of mass energy. 
In \oassx, the running of the strong coupling $\alpha_s $ at 
the renormalization scale $\mu $ is given by 

\begin{equation}
\label{rscale} 
    \alpha_s( \mu ) = 
    \frac {1} { \beta_0 \ln \frac { \mu^2 } { \Lambda^2 } }
    \left(  1 - \frac { \beta_1 } { \beta_0^2 }
                \frac { \ln \ln \frac { \mu^2 } { \Lambda^2 } }
                      { \ln \frac { \mu^2 } { \Lambda^2 } } \right)  
\end{equation} 

\nin
where $\Lambda \equiv \Lambda_{ \overline{MS}}^{(5)} $ is the QCD 
scale parameter computed in the \MSB scheme 
for $n_f = 5 $ flavors and $\beta_1 = (153-19 n_f) / 24 \pi^2 $. 
The renormalization scale $ \mu $ is a formally unphysical parameter
and should not enter at all into an exact infinite order calculation. 
However, within the context of a truncated finite order perturbative 
expansion for any particular process under consideration, the 
definition of $\mu $ depends on the renormalization scheme employed, 
and its value is in principle completely arbitrary. \\


\begin{figure} [tbp]
\begin{center}
\mbox{\epsfig{file=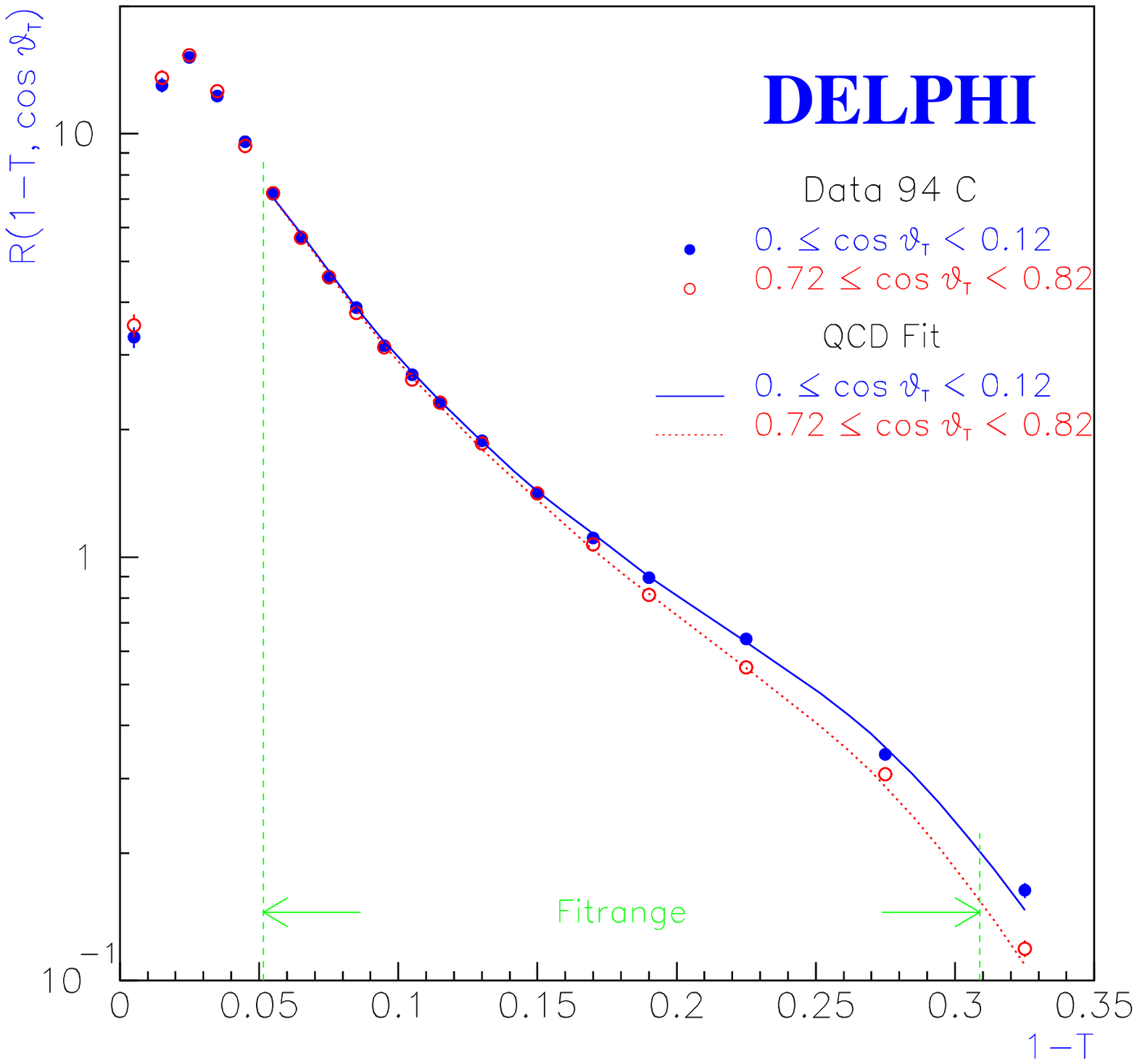, width=7.5cm}}
\mbox{\epsfig{file=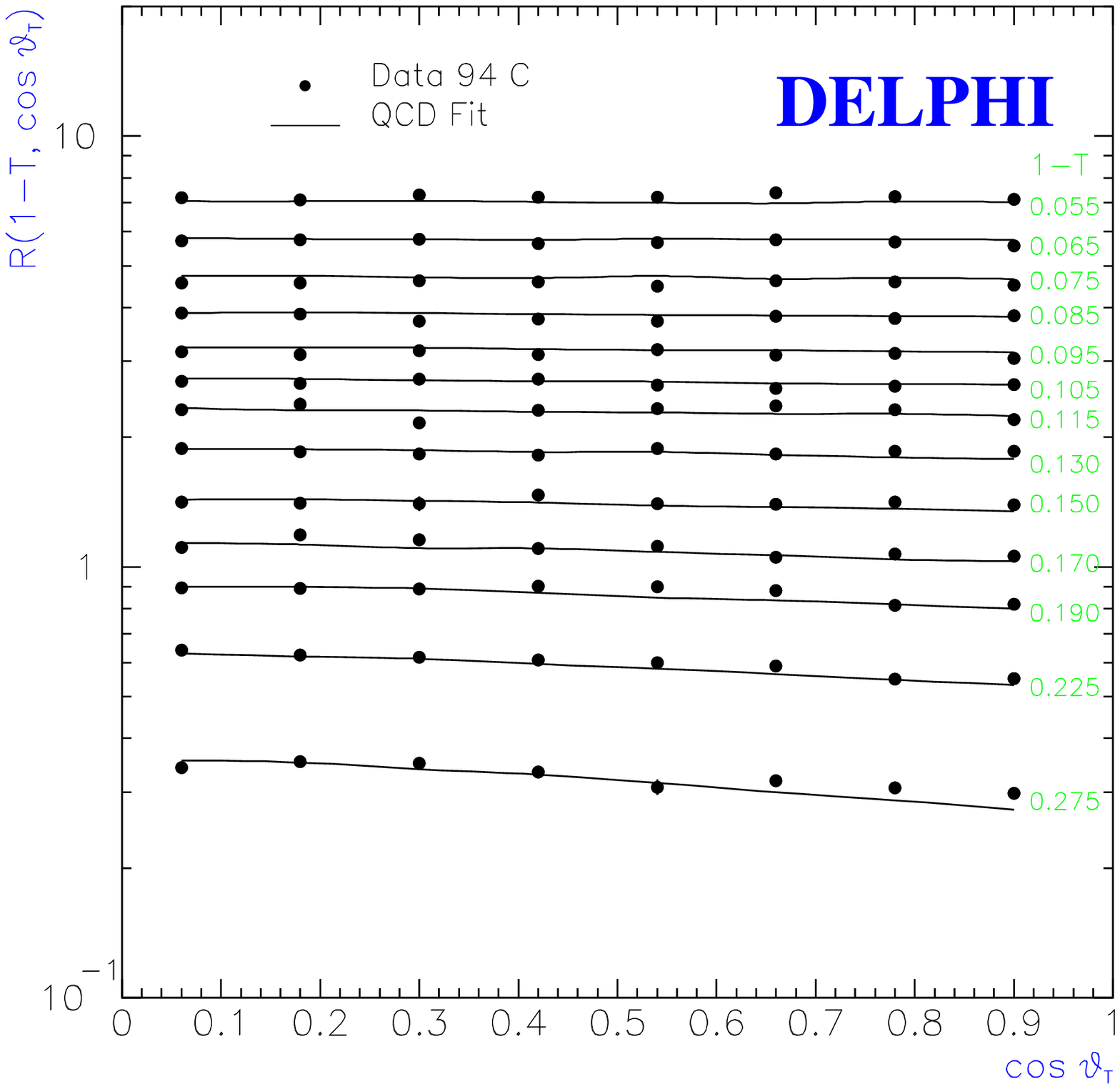, width=7.5cm}}
\end{center}
\caption[]{(a) QCD fits to the measured thrust distribution for two bins
         in $ \rm \cos \vartheta_T $. (b) Measured thrust distribution
         at various fixed values of $\rm 1-T $ as a function of 
         $ \rm \cos \vartheta_T $. The solid lines represent the QCD fit. } 
\label{DataThr}
\end{figure}

\begin{figure} [tb]
\begin{center}
\mbox{\epsfig{file=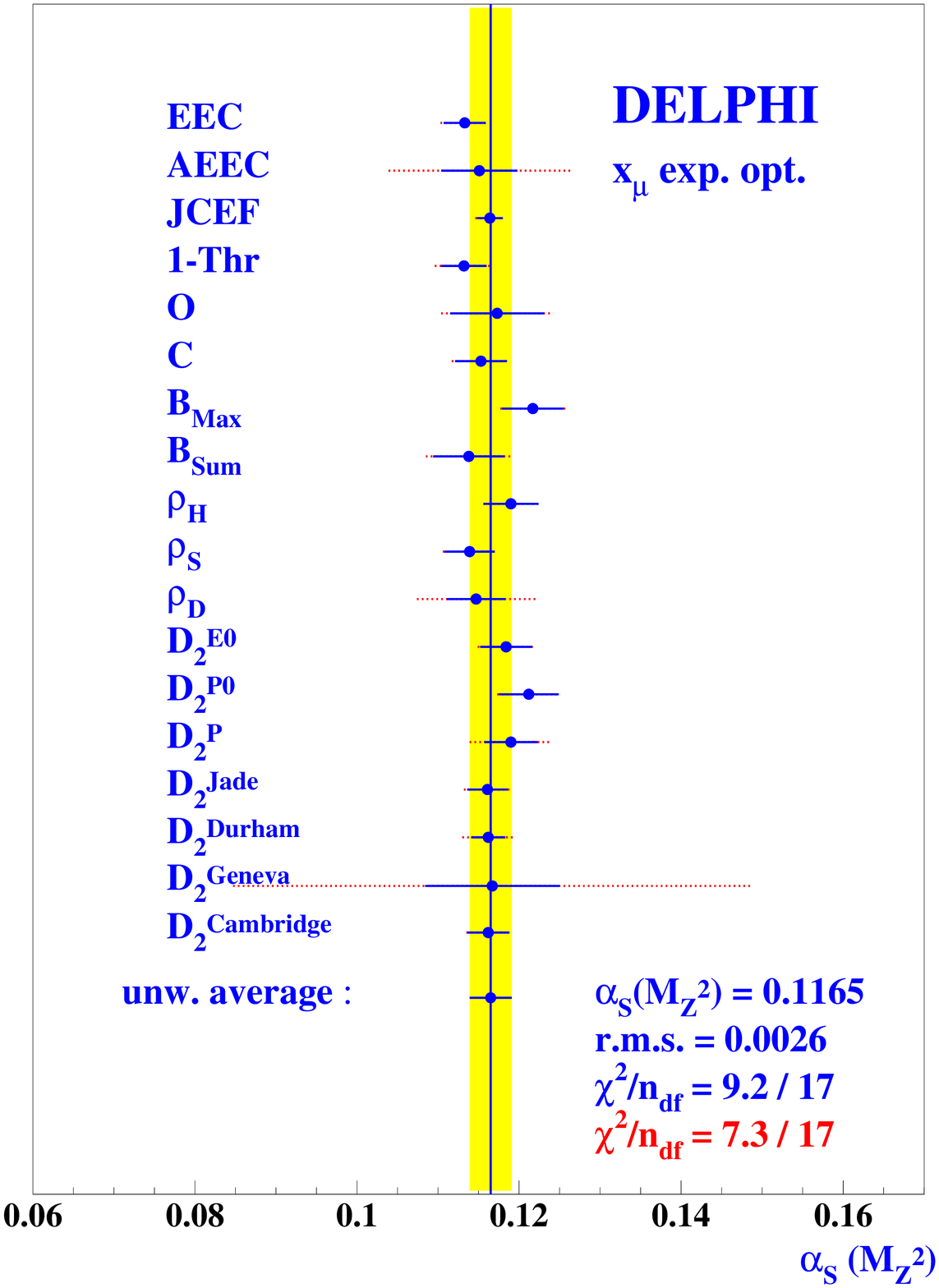, height = 8.5 cm }}
\hspace{0.5 cm}
\mbox{\epsfig{file=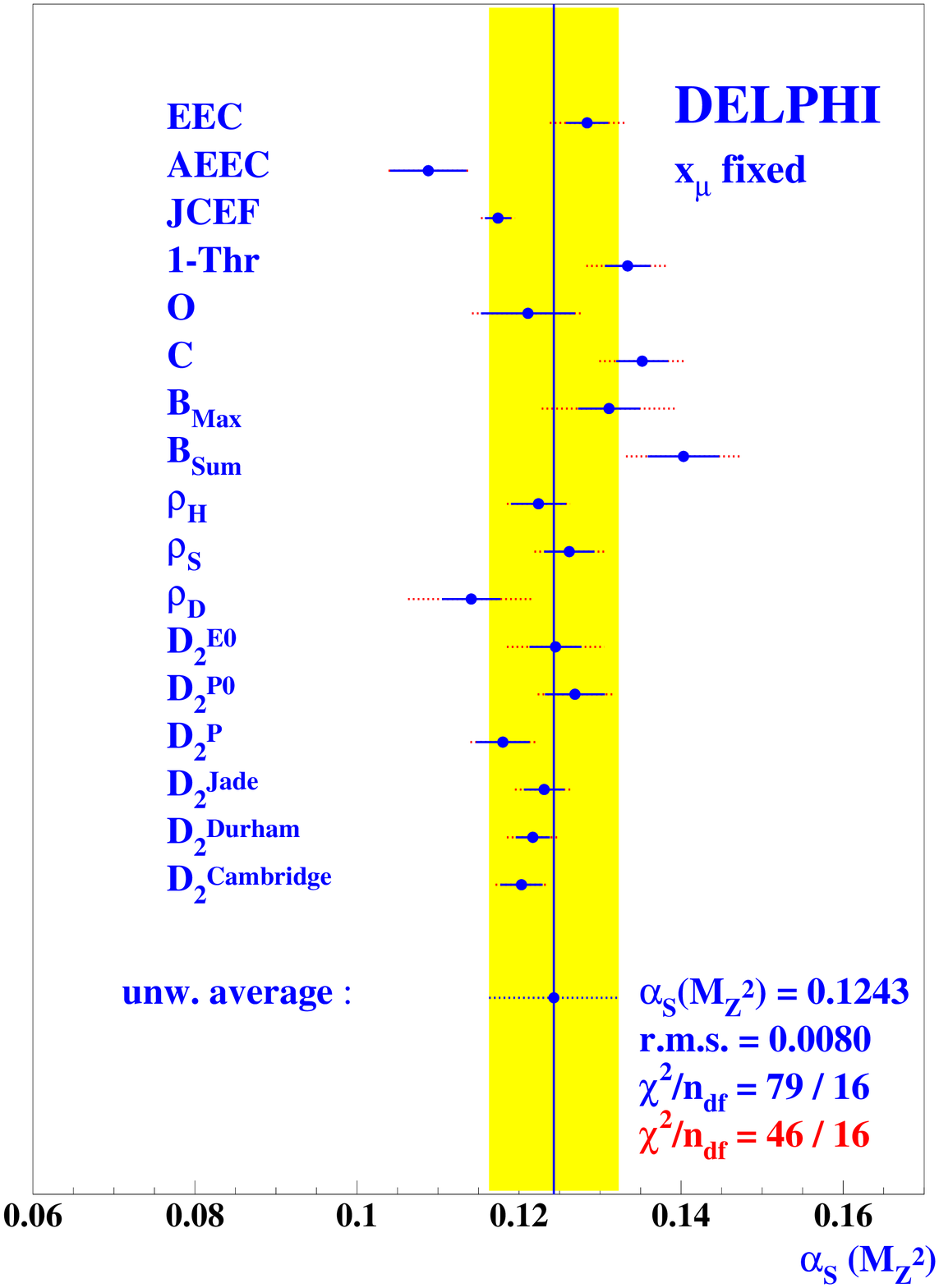, height = 8.5 cm }} 
\end{center}     
\caption[]{Results of the \asmz measurements from 18 event shape 
           distributions. (a) fits using experimentally optimized 
           scale values, (b) fixed scale fits : $ x_{\mu} = 1 $. The 
           errors on \asmz indicated correspond to the quadratic sum 
           of the uncertainty from the fit, the systematic experimental 
           uncertainty and the hadronization uncertainty. The dotted 
           error bars indicate the additional uncertainty due to the 
           variation of the central renormalization scale value 
           $ x_{\mu} $ between $ 0.5 \cdot x_{\mu} $ and 
           $ 2. \cdot x_{\mu} $. }
\label{averuwg}
\end{figure}

The traditional experimental approach is, to measure all observables 
at the same, fixed scale value, the so-called physical scale (PHS) 
$\rm x_{\mu} = 1 $ or equivalently $\rm \mu^2 = Q^2 $. 
Applying PHS to the high precision DELPHI data at $ \sqrt{s} = M_Z $ 
yields in general only a poor description of the measured event 
shape distributions, $ \chi^2 $ values up to 
$ \chi^2 / n_{df} \sim 40 $ are found. For the PHS choice the $\rm 2^{nd} $ 
order contribution in Eq. \ref{Xsect} is in general quite large.
For some observables the ratio of the 
\oass with respect to the \oas contribution $ r_{NLO} $ is almost 
$\rm \left| r_{NLO} \right| \simeq 1.0 $, indicating a poor convergence 
behavior of the corresponding perturbative expression. This quite 
naturally explains the observation of the wide spread of the 
measured \as values for the individual observables 
(see Fig. \ref{averuwg}b). If PHS is applied, an unweighted average 
for the \asmz values of the 18 observables yields  
$ \chi^2 / n_{df} = 40 / 16 $, i.e. the individual measurements are 
inconsistent. For the differential 2-jet rate determined
using the Geneva-Algorithm, the fit applying $ x_{\mu} = 1 $ fails
completely to describe the data. Here, no \asmz value can be derived 
at all if PHS is applied. 
Therefore, the central method for measuring \asmz has been chosen  
to be a combined fit of \as and the scale parameter 
$x_{\mu} $, a method known as experimentally optimized scales (EXP). 
Here one finds in general much smaller contributions
from the \oass term in Eq. \ref{Xsect}, indicating a better convergence 
behavior of the perturbative series. Applying EXP, 
the \oass predictions including the event orientation yield an 
excellent description of the high statistics data 
(see Fig. \ref{DataThr} as an example). For all 
observables considered, the QCD fit yields 
$ \chi^2 / n_{df} \simeq 1 $.   \\


\begin{figure} [tb]
\begin{center}
\mbox{\epsfig{file=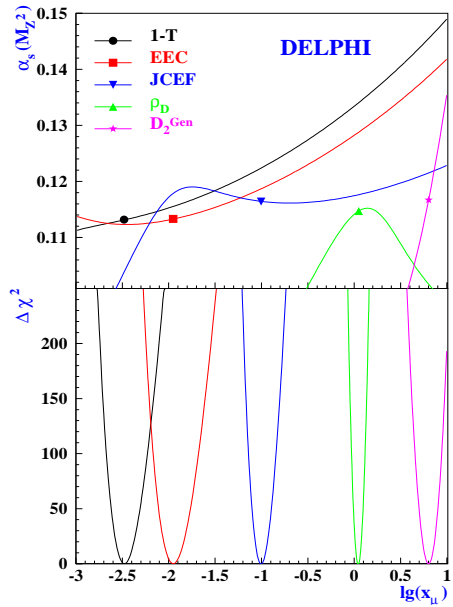, height = 8.0 cm, width = 6 cm }}
\end{center}
\caption[]{\asmz and $\Delta \chi^2 = \chi^2 - \chi^2_{min} $ 
           from \oass fits for some of the observables studied as 
           a function of the renormalization scale $x_{\mu}$. Additionally, 
           the $\chi^2 $ minima are indicated in the \asmz curves. }
\label{scales}
\end{figure}


Fig. \ref{scales} shows the renormalization scale dependence of \asmz 
for some of the observables studied. It turns out that in order to 
describe the data, the scale has to be fixed to a rather narrow range 
of values. Consistent \asmz measurements can only be derived, if the
optimized scale values $\bar{x}_{\mu} $ are applied, i.e. from the \as 
values corresponding to the $ \chi^2 $ minima of the individual fits.
For most of the observables the scale dependence in the vicinity of the 
$ \chi^2 $ minima is significantly reduced, but even for the few 
observables exhibiting a strong scale dependence around the $\chi^2 $ 
minima, the corresponding \asmz values are perfectly consistent.
The observable with the smallest scale dependence of \asmz
is the Jet Cone Energy Fraction (JCEF) \cite{JCEF}. Here, the 
change in \asmz is only \dasmz $ = \pm 0.0010 $, 
even if the scale is varied within the large range of
$\bar{x}_{\mu} / 2 \le x_{\mu} \le 1. $ Additionally JCEF has 
the smallest hadronization correction uncertainties as well.   \\

 The \asmz values determined from 18 different observables are shown
in Fig. \ref{averuwg} for EXP in comparison with PHS.
For EXP the scatter among the different observables is significantly 
reduced. The errors of \asmz correspond to the quadratic sum of the 
uncertainty from the fit, the systematic experimental uncertainty and 
the hadronization uncertainty. Conservatively, an additional uncertainty 
due to the variation of the central renormalization scale value 
$ \bar{x}_{\mu} $ between $ 0.5 \cdot \bar{x}_{\mu} $ and 
$ 2. \cdot \bar{x}_{\mu} $ has been considered for
both methods PHS and EXP. An unweighted average yields 
\asmz $ = 0.1165 \pm 0.0026 $ for EXP to be compared with 
\asmz $ = 0.1243 \pm 0.0080 $ in the case of PHS. The corresponding 
$ \chi^2 $ value is $ \chi^2 / n_{df} = 7.3 / 17 $ for 
EXP and $ \chi^2 / n_{df} = 79 / 16 $ for PHS. It should be 
emphasized, that the consistency for EXP does not depend on the 
additional uncertainty due to renormalization scale variation.
Ignoring this uncertainty yields an consistent average value of \asmz 
as well ($ \chi^2 / n_{df} = 9.2 / 17 $). A weighted average of \asmz
considering correlations between the observables yields 
\asmz $ = 0.1164 \pm 0.0025 $, almost identical to the unweighted 
average. The investigation of the influence of heavy quark mass 
effects on \asmz is under study. A preliminary estimate yields 
\asmz $ = 0.117 \pm 0.003 $. 


\subsection{Theoretically Motivated Scale Setting Methods}


\begin{figure} [tb]
\begin{center}
\mbox{\epsfig{file=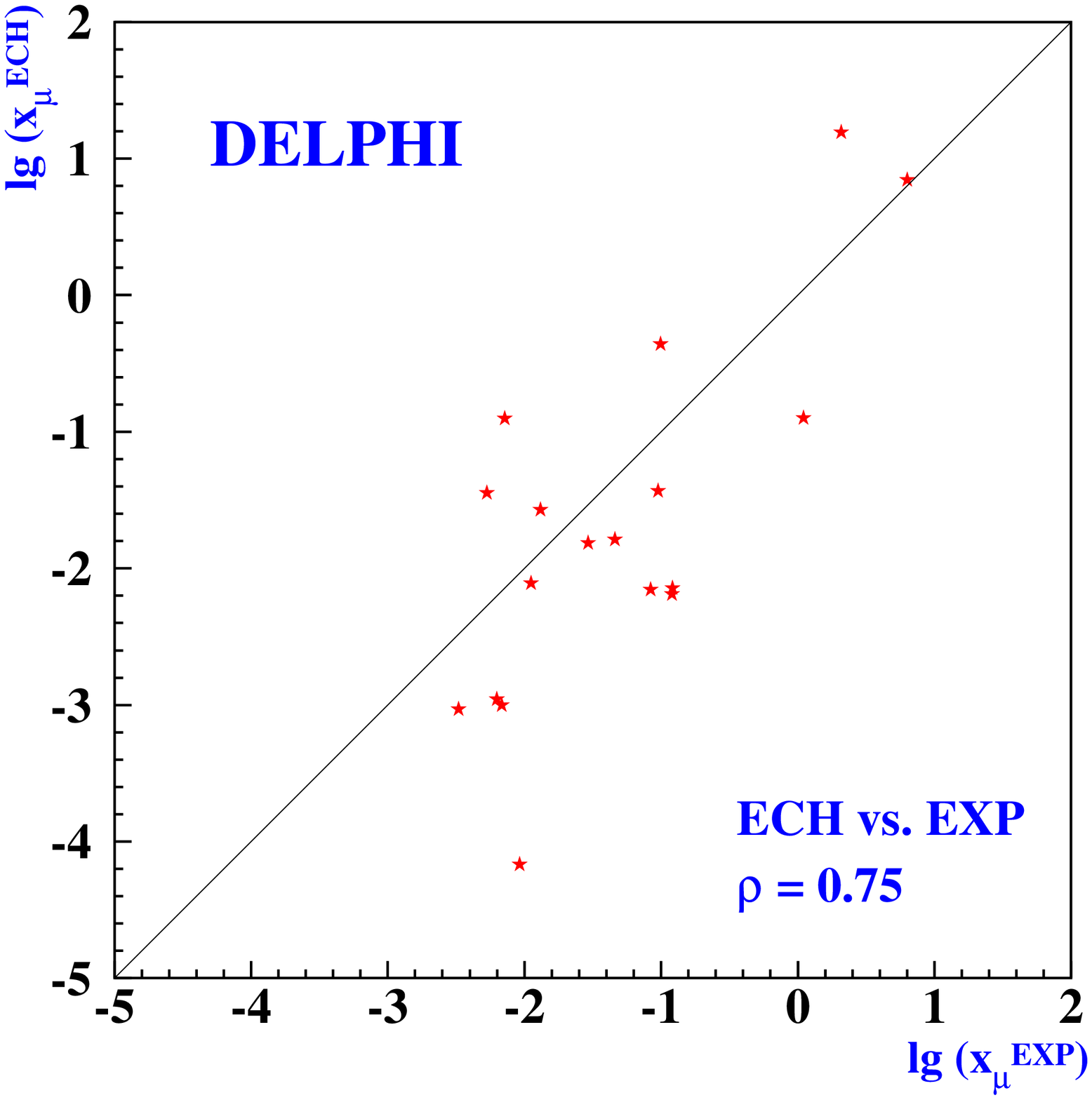, width=7.5 cm}}
\mbox{\epsfig{file=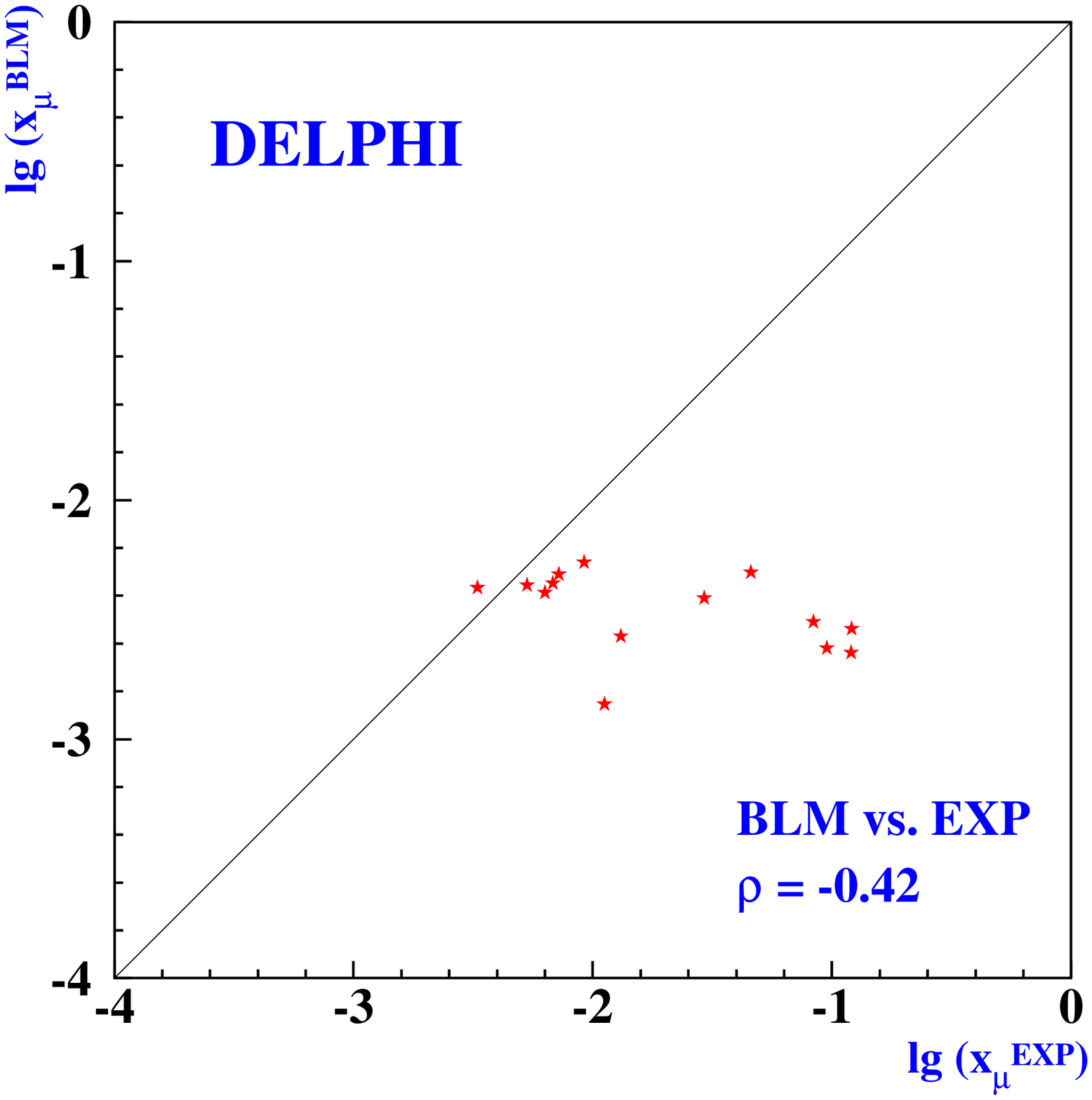, width=7.5 cm}}
\end{center}
\caption[]{Correlation of experimentally optimized renormalization 
           scale values with values predicted by theoretically
           motivated scale setting prescriptions. 
           ({\it left side:} ECH method, 
           {\it right side:} BLM approach)    } 
\label{xcorel}
\end{figure}


Additional studies \cite{delpap} have been made using theoretically 
motivated scale setting prescriptions, like the principle of minimal 
sensitivity (PMS), the method of effective charges (ECH) and the method of 
Brodsky, Lepage and MacKenzie (BLM). In the case of PMS and ECH, a strong 
correlation with the measured scale values from EXP can be observed. 
For the BLM method no such correlation is observed 
(see Fig. \ref{xcorel}). Moreover, the 
BLM fits do not converge for some of the observables under study. 
The individual \as values from the remaining observables turn out 
to be inconsistent. However, the average values of \asmz for all methods 
considered are consistent with EXP, but the scatter of the individual 
\asmz measurements is somewhat larger for the theoretically motivated 
methods.  \\

The influence of higher order contributions has also been investigated 
\cite{delpap} by using the method of Pad\'{e} Approximants for the estimate 
of the uncalculated \oasss contributions (PAP). In comparison with \oassx, 
the renormalization scale dependence for PAP is significantly reduced. 
Here, a fixed scale value of $ x_{\mu} = 1 $ has been chosen for the 
\asmz measurements from the individual observables. The average 
value of \asmz is again in perfect agreement with the \oass result, 
suggesting small contributions due to missing higher order predictions.


\subsection{Comparison with NLLA predictions}

The probably most relevant check on the influence of higher order
contributions comes from a study of the all orders resummed next to
leading logarithmic approximation (NLLA), which has been calculated for
a limited number of observables. Two different strategies
have been applied \cite{delpap}. Pure NLLA calculations have been used to 
measure \asmz in a limited kinematical region close to the infrared limit,
where the logarithmic contributions become large. Matched NLLA + \oass
calculations have been used to extend the \oass fit range towards the 
2-jet region. Unlike \oass theory, no optimization is involved in 
adjusting the renormalization scale for NLLA predictions. 
Therefore the renormalization scale value has been fixed to 
$ x_{\mu} = 1 $. Both methods yield average values of \asmz compatible
with the average from \oassx, the scatter of the individual measurements
is somewhat larger for the resummed theory.    \\


\begin{figure} [tb]
\begin{center}
\mbox{\epsfig{file=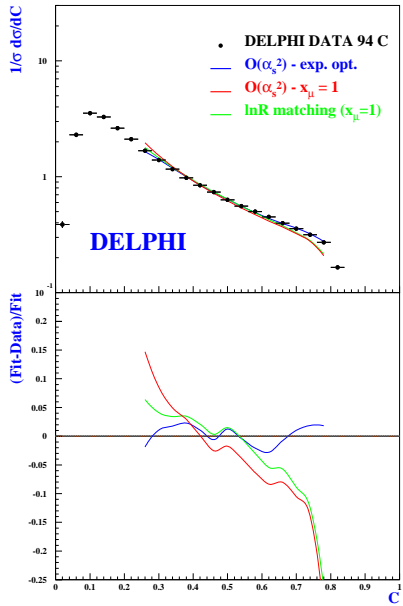, height = 8.5 cm, width = 6 cm}}
\end{center}
\caption[]{ Comparison of the measured C-Parameter Distribution  
            with three different QCD Fits: 
            {\it i})   \oass using an experimentally optimized 
                       renormalization scale, 
            {\it ii})  \oass using a fixed renormalization scale 
                       $\rm x_{\mu} = 1 $ and
            {\it iii}) $ \ln R $ matched NLLA ($\rm x_{\mu}=1 $) 
                       for the C Parameter.
            The lower part shows the relative difference (Fit-Data)/Fit.
            Whereas the \oass curve describes the data over the whole
            fit range, the slope of the curves for the fixed scale 
            and $ \ln R $ matching show a similar systematic distortion 
            with respect to the data. }
\label{lnrvgl}
\end{figure}


Looking at the individual measurements, one finds that the \as values 
derived from matched predictions are systematically higher than those 
from pure NLLA theory. For some of the observables, the \as value from 
matched predictions are even higher than for both pure NLLA and \oass 
predictions. Clearly the matched result is expected to be a
kind of average of the results from both the distinct theories. 
Moreover, the matched theory predictions for some of the 
observables yield only a poor description of the high precision data,
a more detailed investigation reveals a systematic deviation of the 
predicted slope with respect to the data (see Fig. \ref{lnrvgl}). 
The distortion observed is similar
to the distortion obtained using \oass predictions applying a fixed 
renormalization scale value of $ x_{\mu} = 1 $, indicating that the
terms included from the $ 2^{nd} $ order predictions dominate the matched
predictions. Whereas $ x_{\mu} = 1 $ seems to be an appropriate choice
for pure NLLA predictions, the similarity of the two fit curves 
indicate a mismatch of the renormalization scale values for the 
\oass and NLLA part of the combined prediction. 


\section{Other High Precision \bfas Measurements}


\subsection{\hspace{-0.2 cm}\bfas Determination from Precision 
            Electroweak Measurements}

The precise electroweak measurements performed at LEP and SLD can be 
used to check the validity of the Standard Model and to infer information
about its fundamental parameters. Within the Standard Model fits of             
the LEP Electroweak Working Group, the value of \asmz depends mainly
on $R_{l} $, $\Gamma_{Z} $ and $ \sigma_{had} $. The theoretical 
prediction is known in NNLO-Precision. A recent update of the fit 
results presented at the ICHEP'98 \cite{gruen} yields a value for the strong 
coupling of \asmz $ = 0.119 \pm 0.003_{Fit} \pm 0.002_{Theo}$. 


\subsection{\hspace{-0.2 cm}\bfas Determination from Spectral Functions 
            in Hadronic \bftau Decays} 

One of the few processes, where QCD predictions are known in NNLO,
is the hadronic decay of the $\tau $ Lepton. Within the recent OPAL
analysis \cite{opaltau}, \as has been determined from the moments
of the spectral functions of the vector and axial-vector current in
hadronic  $\tau $ decays, which are weighted integrals over the 
differential decay rate $ dR_{\tau,V/A}/ds $ for vector (V) and 
axial-vector (A) decays:


\begin{figure} [tb]
\begin{center}
\mbox{\epsfig{file=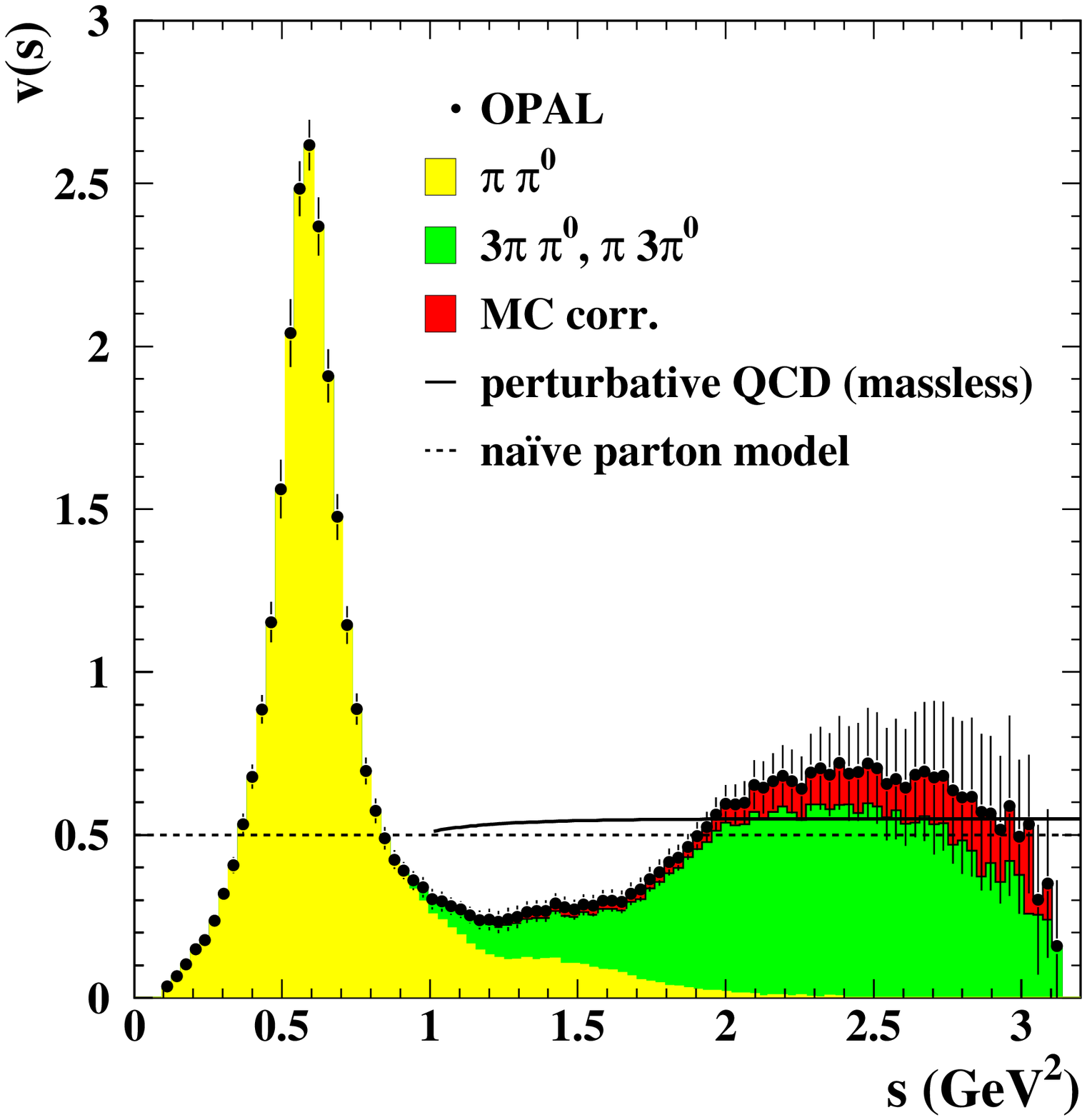, height = 7.0 cm }}
\mbox{\epsfig{file=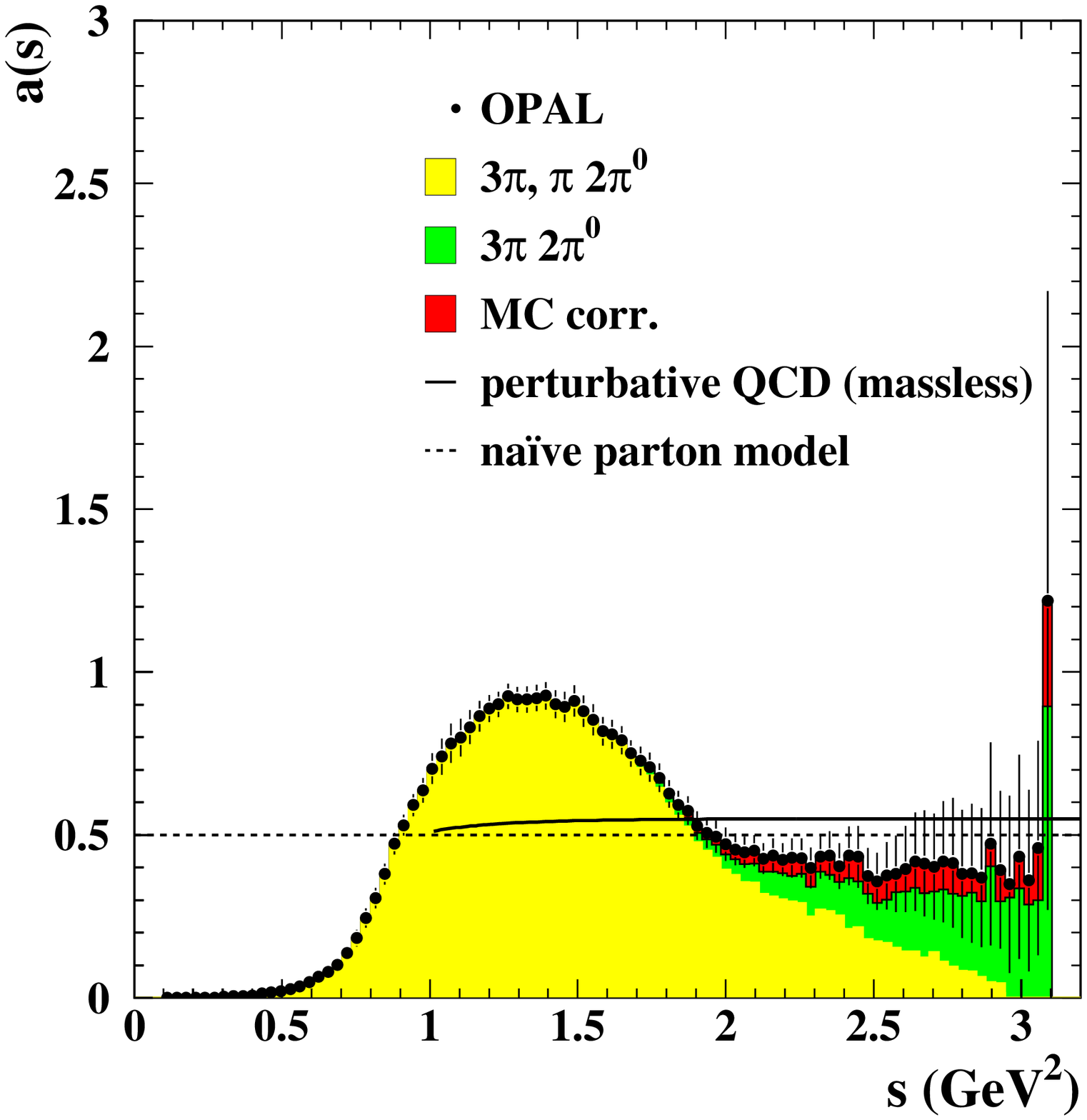, height = 7.0 cm }} 
\end{center}     
\caption[]{The vector and axial-vector spectral functions in hadronic 
           $\tau $decays. The data points correspond to the sum of all
           contributing channels. Some exclusive contributions are 
           shown as shaded areas.}
\label{tauspectra}
\end{figure}


\begin{equation}
\label{taumoments}
      R^{kl}_{\tau,V/A}(s_{0}) = \int\limits_{0}^{s_{0}} ds 
      \left( 1 - \frac{s}{s_{0}} \right)^{k}
      \left( \frac{s}{m_{\tau}^{2}} \right)^{l}
      dR_{\tau,V/A}/ds
\end{equation}

The analysis involves a measurement of
the invariant mass of the hadronic system and thus requires an 
exclusive reconstruction of all hadronic final states. After
unfolding the measured spectra, normalizing them to their
branching ratios and summing them up with their appropriate 
weights, one obtains the spectral functions shown in Fig.
\ref{tauspectra}.    \\

The QCD predictions have been calculated including 
perturbative and non-per\-tur\-ba\-tive contributions. Within 
the framework of the Operator Product Expansion (OPE) the 
non-perturbative contributions are expressed as a power 
series in terms of $ 1/m_{\tau}^{2} $. In contrast to the
perturbative part, the power corrections differ for the vector
and axial-vector part, thus the difference of the vector and 
axial-vector spectral function is sensitive to non-per\-tur\-ba\-tive
effects only. The perturbative contribution is known in \oasss
and partly known in \oaSSx. On a first glance this accuracy 
looks quite impressive, it should however be emphasized, that
with the small momentum transfers involved, the divergence
of the perturbative series is a much more important issue for
the $\tau $ decay than for \as measurements at high energies.
Recent estimates of the complete \oaSS contribution\cite{taupade} to 
the perturbative prediction by the means of Pad\'{e} Approximation
predict an \oaSS contribution of about 20 \% w.r.t. the leading
order term, indicating that further higher-order terms could have
a significant effect on the perturbative prediction.
Three different calculations for the perturbative prediction 
have been studied within the OPAL analysis. Apart from the
standard fixed order perturbative expansion (FOPT), two 
attempts have been studied to obtain a resummation of some
of the higher order terms. The so-called contour improved 
perturbative theory (CIPT) accounts on higher order logarithmic 
terms in \as by expressing the perturbative prediction by 
contour-integrals, which are evaluated numerically using the
solution of the renormalization group equation (RGE) to 
four-loops. The third calculation includes an all order
resummation of renormalon contributions in the so-called  
large $\beta_{0} $-limit (RCPT). This strategy has the advantage
to be renormalization scheme invariant.  \\


\begin{table}
\begin{center}
\begin{tabular}{lccc}         
  \hline
  \multicolumn{4}{l}{Prediction \hspace{2.4 cm}  \asmz  \hspace{2.5 cm} 
                     \chidf \hspace{-0.1 cm}(Fit 1)   
                     \chidf \hspace{-0.1 cm}(Fit 2)                    }  \\      
  \hline
  FOPT       & \hspace{0.2 cm}
               $ 0.1191 \pm 0.0008_{exp} \pm 0.0013_{theo} \pm 0.0003_{evol} $ & 
               \hspace{0.0 cm} $ 0.17 / 1 $ & 
               \hspace{0.3 cm} $ 0.62 / 4 $ \\
  CIPT       & \hspace{0.2 cm}
               $ 0.1219 \pm 0.0010_{exp} \pm 0.0017_{theo} \pm 0.0003_{evol} $ & 
               \hspace{0.0 cm} $ 0.16 / 1 $ & 
               \hspace{0.3 cm} $ 0.63 / 4 $ \\
  RCPT       & \hspace{0.2 cm} 
               $ 0.1169 \pm 0.0007_{exp} \pm 0.0015_{theo} \pm 0.0003_{evol} $ & 
               \hspace{0.0 cm} $ 0.07 / 1 $ & 
               \hspace{0.3 cm} $ 0.61 / 4 $ \\   
  \hline 
\end{tabular}
\end{center}
\caption{Results on \asmz from an OPAL analysis of spectral functions 
         in hadronic $ \tau $ decay. Given are the \asmz values for
         the three different perturbative calculations studied. The
         \asmz values for the two different fits are identical. Shown
         are also the \chidf values for both fit strategies.}
\label{astau}
\end{table}


Two different fits to the data have been performed. The first fit
uses 5 moments from the sum of the vector and axial-vector 
spectral function for the determination of \asmt together with
three parameters from OPE, the second fit uses 10 moments and applies 
the vector and axial-vector functions separately for the 
determination of  \asmt in combination with 5 parameters from OPE.
Both fits yield nearly identical results for \asmtx. The \as values
are extrapolated to $ M_Z $ using the RGE, the results for \asmz
are summarized in table \ref{astau}. The three different approaches
describe the data equally well, the theoretical errors as well as the
overall error is nearly the same, so from an experimental point of view
there is no preferred calculation. However, the difference of the \as
values measured applying different theoretical assumptions is about 4 \%
i.e. the difference is much larger than the the theoretical error 
determined for each method. The reason for this is most likely due to
underestimation of the uncertainty due to missing higher order terms.
For the \asmt determination, the renormalization scale value
$ x_{\mu} = \mu^{2} / m_{\tau}^{2} $ has been fixed to $\rm x_{\mu} = 1 $,
an uncertainty has been estimated due to scale variation in the range
$\rm 0.4 \le x_{\mu} \le 2 $, which is nearly the same range as in the
DELPHI analysis of \ee event shapes. However, within the DELPHI analysis
it turned out, that this scale variation range is sufficient only 
if one applies experimentally optimized scales, but yields inconsistent 
results for fixed scale values. An uncertainty due to the scheme 
dependence of the RGE coefficient $\beta_{3} $, which has been applied 
using the \MSB value, has been estimated due to variation between 
$ 0. \le \beta_{3}^{RS} / \beta_{3}^{\overline{MS}} \le 2.$, however there 
is no theoretical reason, why its value should not be negative. This is
indeed sometimes the case if optimized renormalization schemes are 
applied (se e.g. \cite{freezing}). A similar analysis done by the ALEPH 
collaboration\cite{alephtau} has shown, that the use of the PMS scheme
optimization leads to an reduced value of \asmz and therefore reduces
the discrepancy between FOPT and RCPT. Since the renormalization scheme
optimization turns out to be of major importance in \ee annihilation,
it would clearly be desirable to do similar studies in the analysis
of hadronic $\tau$-decays.   

       
\subsection{\bfas Determination from the Gross-Llewellyn-Smith 
            Sum Rule in $\nu$-N-DIS} 

The Gross-Llewellyn-Smith (GLS) Sum Rule predicts the integral over 
the non-singlet structure function $xF_{3}(x,Q^{2})$ measured in
$\nu$-N deep inelastic scattering. In the naive quark parton model,
the value of this integral should be three. QCD corrections 
have been calculated in NNLO:

\begin{equation}
\label{GLSintegral}
   \int\limits_{0}^{1} xF_{3}(x,Q^{2}) \frac{dx}{x} = 3
   \left[1 - \frac{\alpha_s}{\pi} 
           - a(n_{f}) (\frac{\alpha_s}{\pi})^{2}
           - b(n_{f}) (\frac{\alpha_s}{\pi})^{3}  \right]
   - \frac{\Delta HT}{Q^{2}}   
\end{equation}


\begin{figure} [tb]
\begin{center}
\mbox{\epsfig{file=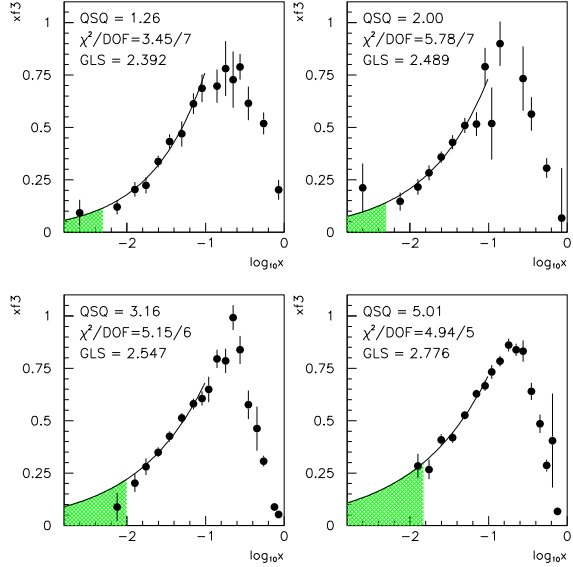, height = 8.0 cm }}
\end{center}
\caption[]{$xF_{3}$ as a function of $x$ at the four lowest 
           $Q^{2}$ values. The curve shows a power law fit to 
           the points with $x<0.1$, which is used to calculate
           the integral in the shaded regions.  }
\label{GLSdata}
\end{figure}


The recent CCFR/NuTeV analysis \cite{GLSsumrule} covers an energy range
of $ 1 \gev^{2} < Q^{2} < 15 \gev^{2} $ with  
$ \langle Q^{2} \rangle \simeq 3 \gev^{2} $. The dominant error comes 
from the contribution of higher twist terms $\Delta HT$, 
which have been estimated to $\Delta HT = 0.15 \pm 0.15 \gev^{2}$.               
The GLS integral is evaluated using the $xF_{3}$ data separately 
in different $Q^{2}$ bins. For very low $x$ the data have been
extrapolated using a power law fit (see fig. \ref{GLSdata}).
Evolving the fit result for $\alpha_s(3 \gev^{2})$ to $M_{Z}^{2}$
yields \asmz = $ 0.114 \pm ^{.005}_{.006}(stat) 
\pm ^{.007}_{.009}(syst) \pm .005(theo)$, which is consistent with 
other precise \as determinations in different energy ranges.


\subsection{\bfas Determination in $\nu$-N-DIS 
            from high $x$ scaling violations}
             
Another \as determination in $\nu$-N-DIS within an updated analysis
of the CCFR collaboration \cite{scalviol} comes from a simultaneous
fit of the $F_{2}$ and $xF_{3}$ structure functions. The QCD 
predictions are known in NLO. Improvements w.r.t. previous analyses
are due to the new energy calibration and the higher energy and 
statistics of the experiment.  The energy range of the data used 
for the \as determination is $ 5 \gev^{2} < Q^{2} < 125 \gev^{2} $. 
The fit yields 
\asmz = $0.119 \pm 0.002(exp) \pm 0.001(HT) \pm 0.004(scale)$.
The precision is better than from the NNLO analysis applying
the GLS sum rule.  \\ 

The result can e.g. be compared with a result of similar precision
from the analysis of the $F_2 $ structure function data from 
SLAC/BCDMS\cite{bcdmsresult}. Within this analysis \as has been 
determined to \asmz = $0.113 \pm 0.003(exp) \pm 0.004(theo)$.
Within both analyses, the renormalization scale value has been
chosen to $x_{\mu} = 1$. For an estimate of the scale uncertainty,
$x_{\mu}$ has been varied within a large range of 
$0.1 \leq x_{\mu} \leq 4.0$. A similar variation has been done
also for the factorization scale uncertainty. The range for
the scale variation has been chosen in such a way, that the $\chi^{2}$
of the fit is not significantly increased, indicating that the
choice $x_{\mu} = 1$ for the central result is appropriate for the
analysis of structure functions in DIS.     


\section{Running of \bfas}

The running of the strong coupling \as is among the most
fundamental predictions of QCD. Due to the large energy 
range covered by LEP1/2, this prediction can now be tested
from \ee data, measured within a single experiment.  
Whereas the perturbative predictions lead to an 
approximately logarithmic energy dependence of event shapes,
the hadronization process causes an inverse power law behavior
in energy and can therefore be disentangled from the 
perturbative part by studying the energy evolution of event shapes. 
Traditionally, hadronization corrections are calculated by 
phenomenological Monte Carlo (MC) models, whose predictions can now 
precisely been tested over a large energy range. Although 
MC-models yield a good description of the measured data 
within a large energy range (see e.g. \cite{delrunning}), 
their predictive power suffers from a large number of free 
parameters, which have to be tuned to the data (e.g. 10-15 
main parameters to be tuned within the JETSET partonshower model).
A novel way in understanding the hadronization process has been 
achieved with the Dokshitzer-Webber (DW) model\cite{DWmodel}, 
which has recently been improved by results from 2-loop-calculations.
Within this model, the power behaving contributions can be calculated, 
leaving only a single non-perturbative parameter to be determined 
from the data.

  
\subsection{Power Corrections and the Dokshitzer-Webber model}
\label{Dwtheo}

Non-perturbative power corrections in the spirit of the DW-model
arise from soft gluon radiation at energies of the order of the
confinement scale. The leading power behavior is quantified by a
non-perturbative parameter $\alpha_0$, defined by

\begin{equation}
\label{alphazero}
   \alpha_0(\mu_{I}) = \frac{1}{\mu_{I}} 
                       \int\limits_{0}^{\mu_{I}}
                       \alpha_s(k)dk 
\end{equation}

\nin
which is expected to be approximately universal.
Here, the true coupling $\alpha_s(k)$ is assumed to be infrared regular
and can be understood as the sum of two terms

\begin{equation}
\label{alphasplit}
   \alpha_s(k) = \alpha_s^{PT}(k) + \alpha_s^{NP}(k)
\end{equation}

\nin  
where the perturbative part $\alpha_s^{PT}$ is separated from the
non-perturbative part $\alpha_s^{NP} $ by an infrared matching 
scale $\mu_{I}$ of the order of a few $ \gev $. It is expected, that
the factorial growing divergence of the fixed order perturbative 
expansion gets cancelled due to the merging with the non-perturbative 
counter-part, yielding a renormalon free theory.  \\

For the shape observables $Y$ studied so far, the DW-model predicts the 
effect of the power corrections to be a simple shift 
of the perturbative distribution, i.e. for $Y = 1-T, C, M_{H}$:
  
\begin{equation}
\label{Pshift}
   \frac{d\sigma}{dY}(Y) = 
   \frac{d\sigma^{pert.}}{dY}(Y-c_{Y}{\cal P})
\end{equation}

\nin
with an observable dependent factor $c_{Y}$ and $\cal P$ 
proportional $1/Q$:

\begin{equation}
\label{Psize}
   {\cal P} = \frac{4 C_{F}}{\pi^{2}}{\cal M} \frac{\mu}{Q}
   \left[ \alpha_0(\mu_{I})-\alpha_s(\mu)-\frac{\beta_{0}}{2\pi}
          \left( ln\frac{\mu}{\mu_{I}} + \frac{K}{\beta_{0}} + 1 \right)
   \right]
\end{equation}

\nin
For the jet broadening observables $Y = B_{W}, B_{T} $ the shift is
predicted to be 

\begin{equation}
\label{Pshiftprime}
   \frac{d\sigma}{dY}(Y) = 
   \frac{d\sigma^{pert.}}{dY}(Y-c_{Y}{\cal P}ln\frac{B}{B_{0}})
\end{equation}

\nin
with an additional non-perturbative parameter $\alpha^{\prime}_{0} $ 
and a term ${\cal P}^{\prime}$ entering via the log-enhanced term. 
Within the re-analysis of the JADE-data\cite{pedro} it turned out 
that in order to describe the transition from the perturbative 
predictions to the observed spectra of the jet broadening observables, 
not only a shift, but also a squeeze of the partonic distribution 
is required. The original calculations for the jet broadening 
observables are considered to be erroneous, however
the problem seems to be solved now\cite{doktalk}.    


\subsection{Power Corrections to mean event shapes}


\begin{figure} [tb]
\begin{center}
\mbox{\epsfig{file=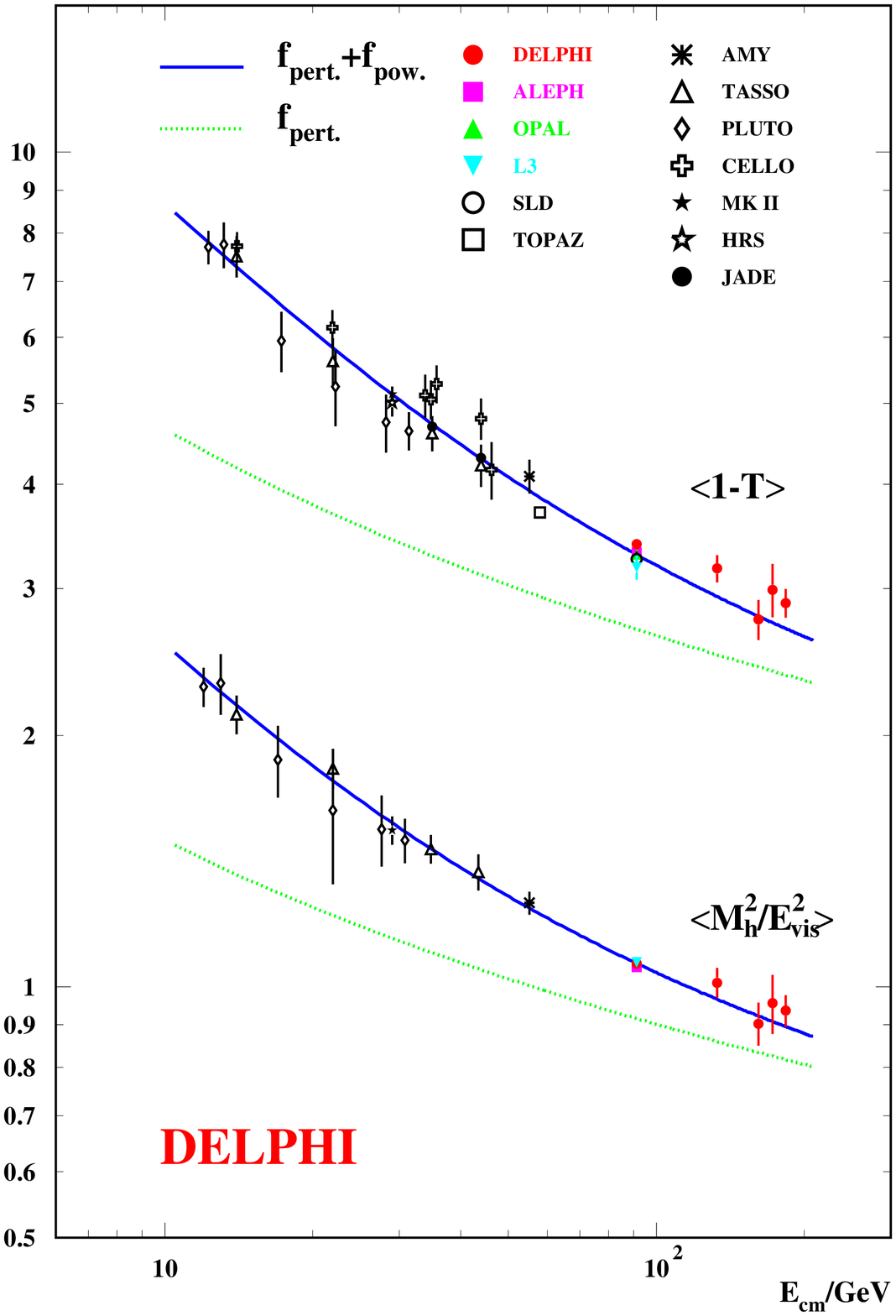, height = 8.5 cm, width = 7.0 cm }}
\mbox{\epsfig{file=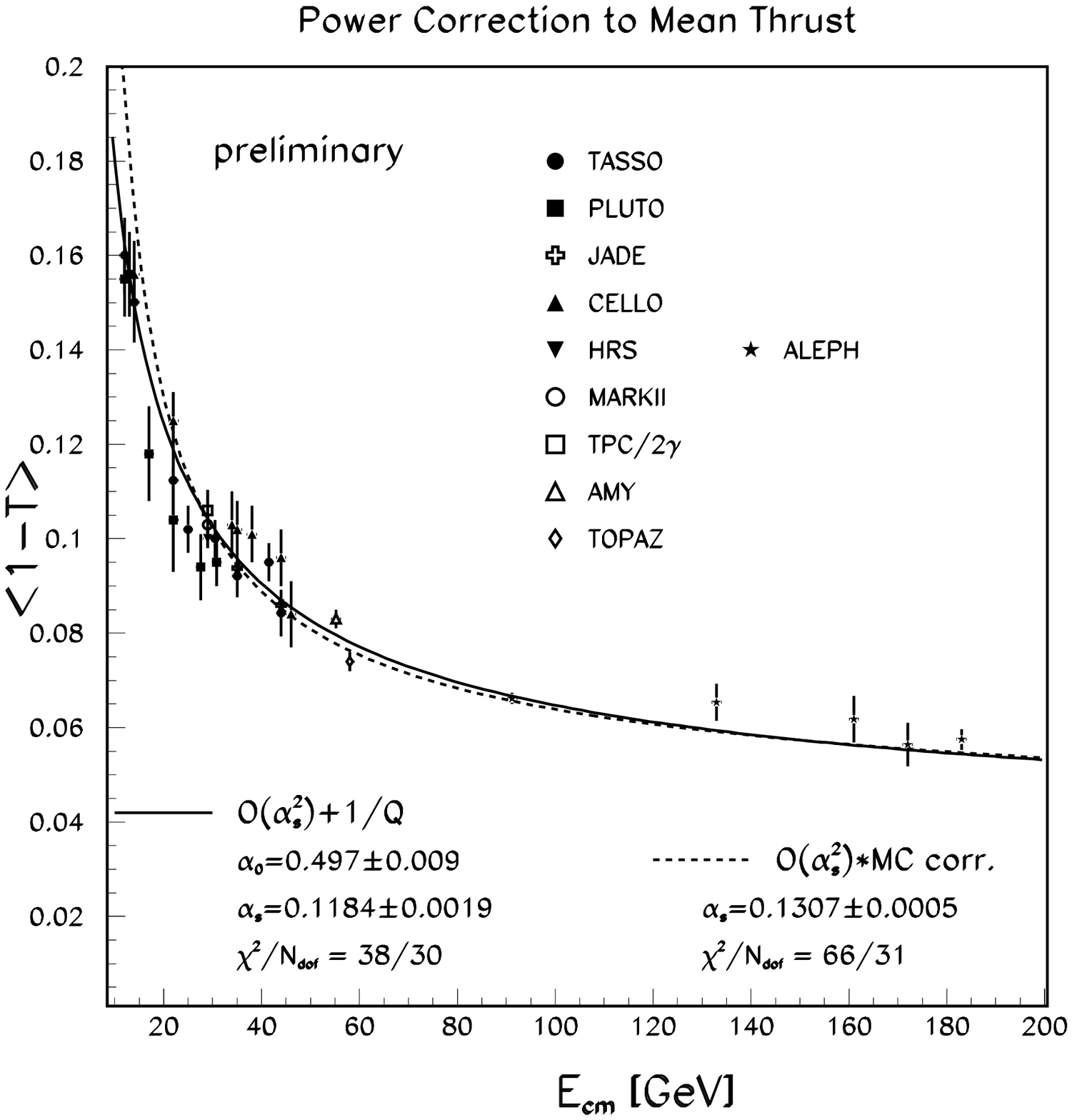, height = 8.0 cm, width = 7.0 cm }} 
\end{center}     
\caption[]{ {\it left side:} Measured mean values of $1-T$ and $M_H$ 
            from DELPHI high energy data and low energy data from 
            various experiments. The solid lines represent the QCD fits 
            in \oass applying the DW-model, the dotted lines show the 
            perturbative contributions only.
            {\it right side:} $\langle 1-T \rangle$ distribution from
            an ALEPH analysis. The solid line represents a QCD fit 
            in \oass applying the DW-model the dashed line represents 
            a fit applying MC hadronization corrections. }
\label{dwshapemeans}
\end{figure}



\begin{table} [b]
\begin{center}
  \begin{tabular}{lcllr}
  \hline
      Analysis                             & 
      Observable                           & 
      \hspace{0.1 cm} $\alpha_0(2 \gev) $  & 
      \hspace{0.4cm}  \asmz                & 
      $ \chi^{2}/n_{df} $                   \\
  \hline
      DW-model (DELPHI)                    & 
      $\langle 1-T \rangle$                & 
      $ 0.494 \pm 0.009 $                  & 
      $ 0.1176 \pm 0.0057 $                &  
      43/22 \hspace{0.0 cm}                  \\
      DW-model (DELPHI)                    & 
      $\langle M_H \rangle$                & 
      $ 0.558 \pm 0.025 $                  & 
      $ 0.1172 \pm 0.0037 $                & 
      2/11 \hspace{0.0 cm}                   \\
      DW-model (ALEPH)                     & 
      $\langle 1-T \rangle$                & 
      $ 0.497 \pm 0.009^{*} $              & 
      $ 0.1184 \pm 0.0019^{*} $            & 
      28/30 \hspace{0.0 cm}                  \\
      MC-corr.\hspace{0.22 cm} (ALEPH)     & 
      $\langle 1-T \rangle$                & 
      \hspace{0.9 cm} $ - $                & 
      $ 0.1307 \pm 0.0005^{*} $            & 
      66/31 \hspace{0.0 cm}                  \\ 
  \hline
  \end{tabular}
\end{center}
\caption{Results of QCD fits in \oass to the mean values of event 
         shapes at different $E_{cm}$. Hadronization corrections 
         have been applied either by the means of the DW-model or 
         as predicted by MC. $^{*}$Statistical errors only.}
\label{shapemeanresults}
\end{table}


The earliest predictions from the DW-model have been made 
for the mean values of event shape distributions. 
In the context of the analysis of LEP2 
data they have the advantage to make use of the full data statistics. 
Fig. \ref{dwshapemeans} shows QCD fits in \oass applying the DW-model 
to mean event shapes measured at various $ E_{cm}$, done by 
DELPHI\cite{delrunning} and ALEPH\cite{alephrunning}.  
The fit results are listed in table \ref{shapemeanresults}. For the
two observables studied, universality of $\alpha_0$ is found on 
a 10 \% level. Apart from a somewhat poor fit quality for the 
$\langle 1-T \rangle$ data used by the DELPHI collab., which can be 
explained by the poor quality of the low energy data, the 
DW-model fits yield a good description of the data, which is even 
better than for MC hadronization corrections.   
The \asmz values obtained applying the DW model are
in remarkable good agreement with \asmz as determined by the DELPHI 
analysis of \ee event shapes at the \ZZx. There is however a
fundamental difference between both analyses: Whereas the analysis
of \ZZ data applies experimentally optimized scales in combination
with MC hadronization corrections, the analysis applying the 
DW-model applies fixed scale values $x_{\mu}=1$, since a scale 
optimization is not feasible in this case. However, the fit quality
of the DW-model fits indicate, that  $x_{\mu}=1$ is an appropriate choice.
The ALEPH result for the QCD fit to the $\langle 1-T \rangle$ 
distribution applying MC hadronization corrections is also 
listed in table \ref{shapemeanresults}. Here, again $x_{\mu}=1$ has 
been applied. The fit quality is quite poor and \asmz is much larger.  
The same observation has been made within the DELPHI \ZZ
analysis, and the result is just another demonstration, that fixed
scale values are not qualified for an analysis applying MC corrections.


\subsection{Power Corrections to event shape distributions}


\begin{figure} [tb]
\begin{center}
\mbox{\epsfig{file=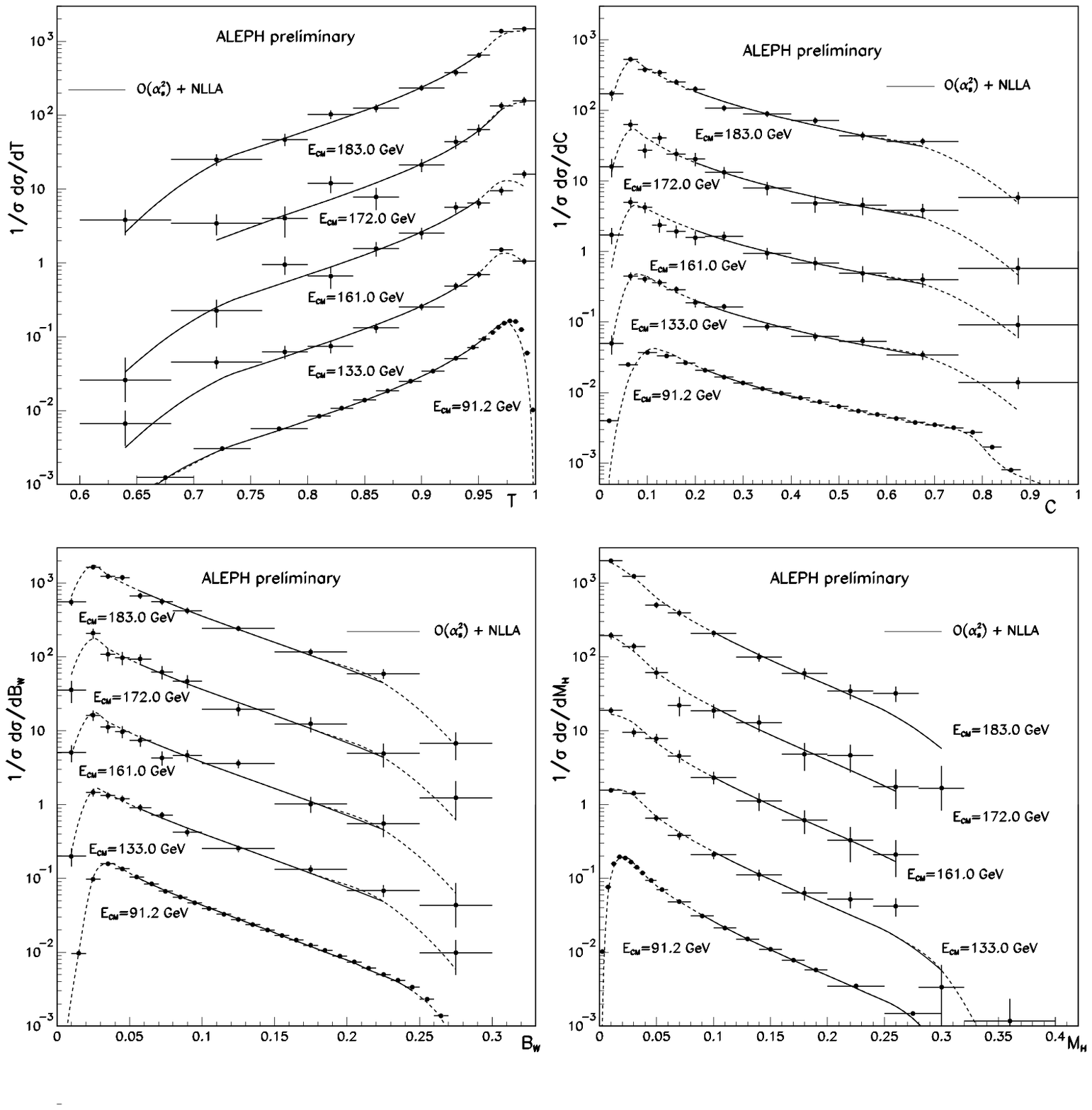, height = 8.0 cm }}
\mbox{\epsfig{file=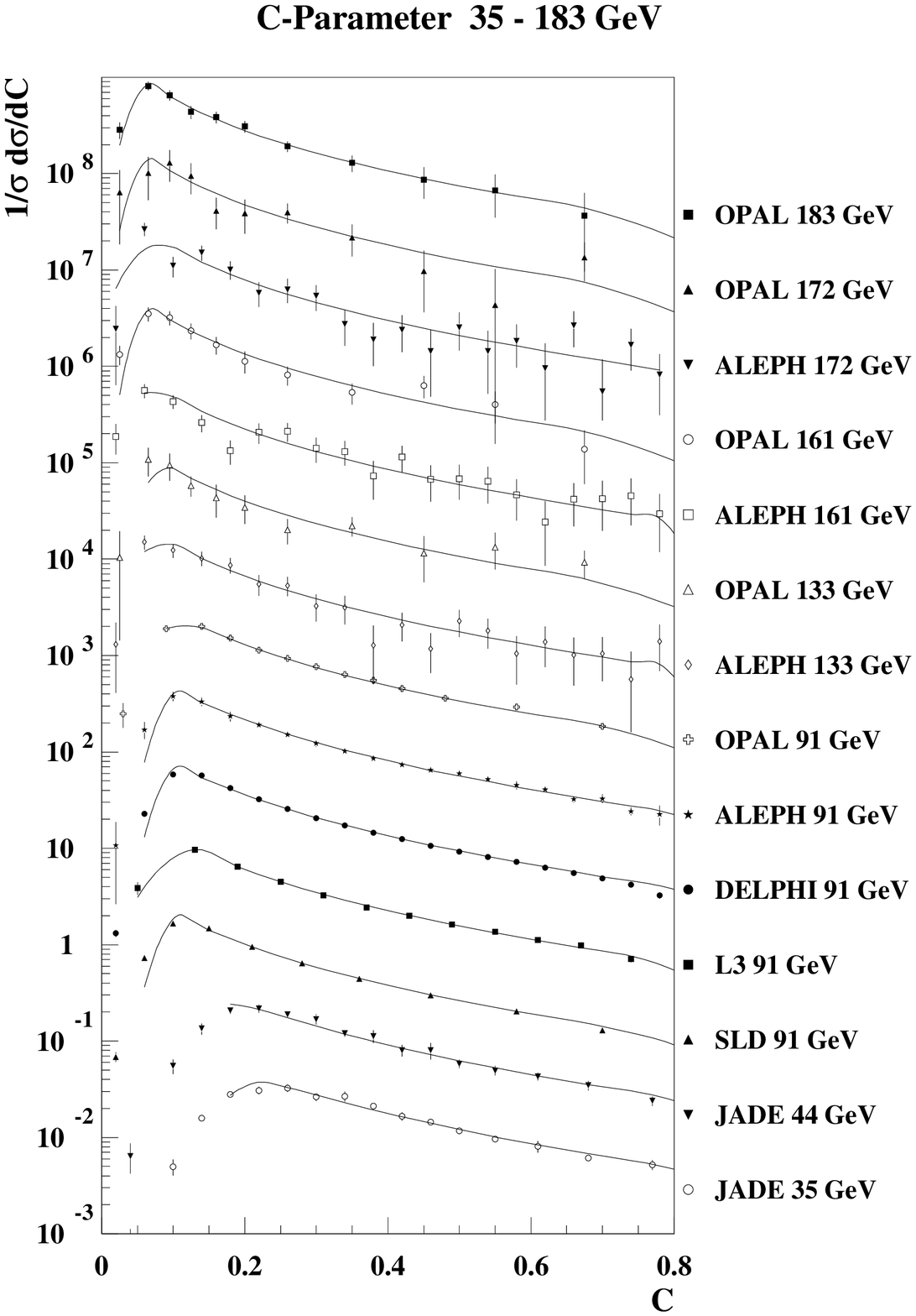, height = 8.0 cm }}
\end{center}
\caption[]{ {\it left side:}
            Measured distributions of Thrust, C-Parameter, wide
            jet broadening and the heavy jet mass from 
            ALEPH data between $91.2 \gev \leq E_{cm} \leq 183 \gev $.
            The lines represent a QCD fit applying the DW-model to 
            shape distributions in matched \oass and NLLA precision.
            {\it right side:}
            QCD fits to the C-Parameter in matched \oass and NLLA 
            precision applying DW power corrections to JADE and LEP
            data between $ 35 \gev \leq E_{cm} \leq 183 \gev$.    }
\label{alephdwshapes}
\end{figure}


ALEPH\cite{alephrunning} has studied event shape distributions of
Thrust, C-Parameter, wide jet broadening and the heavy jet mass in 
matched \oass and NLLA precision. The central results are quoted as
averages between two different matching schemes. They studied as 
well power corrections as hadronization corrections from MC. 
Fig. \ref{alephdwshapes} shows the distributions in the energy 
range $91.2 \gev \leq E_{cm} \leq 183 \gev $. For the fits applying
the DW-model they reported a poor fit quality for the jet 
broadening distribution and that no \as determination was possible. 
This can be understood due to the erroneous theoretical calculation 
mentioned before. For the heavy jet mass they got reasonable results
applying the ln R matching scheme but the fits were unstable under
systematic variations of the analysis. The results for the Thrust and
the C-Parameter fits are listed in table \ref{ALEPHresults}. \\ 


\begin{table}
\begin{center}
\begin{tabular}{lcccc}
\hline
  Analysis        & Observable & $\alpha_0(2 \gev ) $ & \asmz \\
\hline
                  & T         & $ 0.462 \pm 0.060 $ & $ 0.1193 \pm 0.0064 $ \\
    DW-model      & C         & $ 0.449 \pm 0.082 $ & $ 0.1130 \pm 0.0046 $ \\
                  & C,T comb. & $ 0.451 \pm 0.061 $ & $ 0.1168 \pm 0.0044 $ \\
\hline
    MC-correction & T         & $ - $               & $ 0.1253 \pm 0.0061 $ \\ 
    (LEP 1)       & C         & $ - $               & $ 0.1212 \pm 0.0065 $ \\ 
\hline
    MC-correction & T         & $ - $               & $ 0.1282 \pm 0.0054 $ \\ 
    (LEP 2)       & C         & $ - $               & $ 0.1233 \pm 0.0060 $ \\ 
\hline
\end{tabular}
\end{center}
\caption{Results for the ALEPH fits to Thrust and C-Parameter. A combined
         fit to Thrust and C-Parameter has been made for the DW-model 
         analysis. The fits applying MC hadronization corrections have 
         been done separately for LEP1 and LEP2 energies.}
\label{ALEPHresults}
\end{table}



\begin{table} [b]
\begin{center}
\begin{tabular}{lccc}
\hline
   Observable & $\alpha_0(2 \gev ) $ & \asmz \\
\hline
   T         & $ 0.501 \pm 0.009 $ & $ 0.1136 \pm 0.0015 $ \\
   C         & $ 0.482 \pm 0.008 $ & $ 0.1128 \pm 0.0022 $ \\

\hline
\end{tabular}
\end{center}
\caption{Results from QCD fits in matched \oass and NLLA precision 
         applying power corrections predicted by the DW-model to 
         the distributions of Thrust and C-parameter. The errors
         stated are statistical errors only.}
\label{JADEresults}
\end{table}


As for the DW fits to mean event shapes, the quality of the fits to
event shape distributions is better if the DW-model is applied than 
for fits applying MC corrections.  
The agreement with DW-model fits to mean event shapes in \oass is good, 
in particular for the combined fit to the Thrust and C-Parameter.
It should be emphasized, that the observed agreement is different
than the observation within the DELPHI analysis of \ZZ data,
where it turned out, that apart from a poor description of the data,
the \as values from matched predictions were systematically higher 
than for \oass predictions. The results from the DELPHI analysis
suggested a mismatch between the quite different renormalization
scale values required for NLLA and \oass theory. As shown in the
previous section, a fixed renormalization scale value $x_{\mu} = 1$ 
is appropriate if the DW-model predictions are applied, therefore
there should be no mismatch if the matched predictions
are applied in combination with the DW-model. The agreement between
\oass and matched results can then be interpreted in such a way
that the contribution of the higher order logarithmic terms is quite
small. In contrast the \as values from fits applying MC hadronization
corrections are quite large, they are indeed larger than in any of the
high precision analyses introduced before. This observation is 
basically the same than in \cite{delpap}, also the fit quality 
is worse than for the DW-model, but acceptable within this analysis. \\
  
DW-model fits to event shape distributions have also been made within
the re-analysis of the JADE data\cite{pedro} between $ 35 \gev $  
and $44 \gev $
in combination with various data from LEP experiments. Since the 
predictions for the jet broadening observables turned out to be 
erroneous, the results for the fits to Thrust and C-Parameter only
are given in table \ref{JADEresults}. So far, only the statistical 
errors have been evaluated. The results are in agreement with the 
results from the ALEPH collaboration. There seems to be a
trend that the measured \as values are even smaller than \as from
DW-model fits in \oassx, contrary to \as measurements applying 
MC corrections.


\subsection{\bfas and its running from the higher moments 
            of shape distributions}  


\begin{figure} [bt]
\begin{center}
\mbox{\epsfig{file=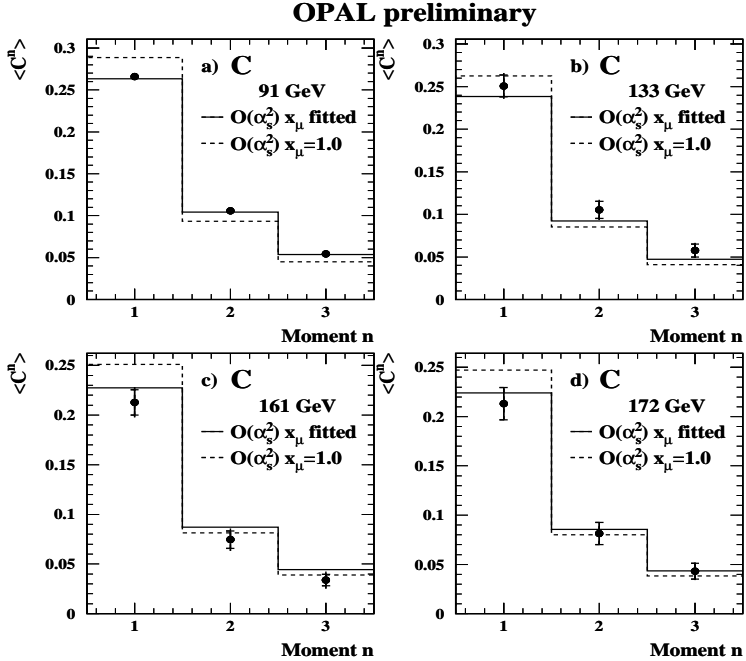, height = 9.0 cm, width = 10 cm }}
\end{center}
\caption[]{Moments of the C-Parameter Distribution at different $E_{cm}$
           from an OPAL analysis. Superimposed is a fit of \oass QCD
           including DW power corrections for the first moment and 
           MC-corrections for the higher moments. The solid line shows
           a fit with experimentally optimized renormalization scales,
           the dashed line with $x_{\mu}=1$ kept fixed.}
\label{opalmoments}
\end{figure}


Even at LEP energies, perturbative QCD predictions yield a significant
correction due to large non-perturbative, power suppressed corrections.
This contributions can be reduced by the study of the higher moments $n$ 
of event shape distributions:

\begin{equation}
\label{moments}
  \langle Y^{n} \rangle = \frac{1}{\sigma} 
  \int dy \frac{d\sigma}{dY} Y^{n}
\end{equation}


\begin{table}
\begin{center}
\begin{tabular}{cc|cc|cc}
\hline
  Observable & Moments & \asmz & $\chi^{2}/n_{df}$ & 
                         \asmz & $\chi^{2}/n_{df}$             \\
             &         & \multicolumn{2}{c|}{$x_{\mu}$ fitted}  
                       & \multicolumn{2}{c}{$x_{\mu}=1$, fixed} \\
\hline
  C          & 1-3     & $ 0.116 \pm 0.010 $ & 9.4/9     & 
                         $ 0.141 \pm 0.010 $ & {\it "large"} \\
  1-T        & 1-3     & $ 0.114 \pm 0.009 $ & 5.8/9     & 
                         $ 0.140 \pm 0.009 $ & {\it "large"} \\
\hline
  C          & 1       & 0.1164 & 2.3/3 & 0.1307 & 2.8/3 \\
  C          & 2       & 0.1164 & 2.7/3 & 0.1537 & 3.1/3 \\
  C          & 3       & 0.1164 & 2.9/3 & 0.1609 & 3.1/3 \\ 
\hline
\end{tabular}
\end{center}
\caption{Fit results of QCD fits in \oass to the moments of event shape
         distributions at energies between 
         $ 91.2 \gev \leq E_{cm} \leq 172 \gev $. DW power corrections 
         have been applied to the first moment only. Results on \asmz 
         are shown for simultaneous fits to the first three moments of
         1-T and the C-Parameter applying experimentally optimized 
         scales and for fits applying a constant renormalization scale 
         value $x_{\mu}\equiv\mu/Q=1$. For the C-Parameter, also the
         \as values for separated fits to the individual higher moments
         are given. For these fits, the uncertainties on \asmz are  
         not available so far.}            
\label{opalmomres}
\end{table}   


\nin
which has recently been performed within an OPAL analysis\cite{opalrunning}. 
For shape distributions with power corrections proportional $\mu_{i}/Q$, 
the power corrections of the corresponding higher moments are expected to be 
suppressed by factors $(\mu_{i}/Q)^{n}$. The size of the power corrections
could in principal be determined by applying the DW-model, in practice
however it turns out, that the corresponding non-perturbative parameter 
$\alpha_{n-1} $ can only be constrained by the data for the first moment
$n=1$. Therefore DW power corrections have only been applied for the first
moments and MC corrections for the second and third moment of thrust and
C-Parameter. QCD fits in \oass have been done to the first three moments
of event shape distributions simultaneously at various $E_{cm}$, 
Fig. \ref{opalmoments} shows for example the moments of the C-Parameter.
The results of the determination of \asmz are listed in 
table \ref{opalmomres}. As for the analysis applying MC corrections 
introduced before, also OPAL finds only a poor description of the data, 
if the renormalization scale value is fixed to $x_{\mu}=1$, 
whereas the description is perfect in the case of experimentally 
optimized scales. In the case of experimentally optimized scales, the 
fitted values of \asmz are in good agreement with the results of the
determination of \asmz in \oass at the \ZZx \cite{delpap} as well as 
with the \asmz determinations applying DW power corrections in \oass 
and matched \oass with NLLA. The largest contribution to the total
uncertainty of \asmz is due to the variation of the renormalization 
scale. The quite conservative estimate of this uncertainty 
yields \dasmz $ = 0.0074(scale)$ for both observables. For the fits
applying a fixed scale value, the \asmz values obtained are however 
much larger, and even under consideration of large scale uncertainties
only poorly compatible with the precise \asmz determinations introduced 
before. The differences of the results between experimentally optimized 
scales and fixed scale values are even more obvious, if one looks at the
results for the fits to the individual moments of the shape 
distributions. They are listed for the C-Parameter as an example in 
table \ref{opalmomres}. Whereas in the case of experimentally optimized
scales \asmz is exactly identical for the separate fits to the first 
three moments, the scatter of \asmz is about 20 \% for the three fits.
The largest value of \asmz = 0.1609 is obtained from a fit to the third
moment, which differs from the current PDG-average by several $\sigma$.
A model with constant \as has been excluded within the OPAL analysis 
with a confidence level of at least 95 \%.


\subsection{Running of \bfas from LEP data between 30 and 183 $\gev$}  


\begin{figure} 
\begin{center}
\mbox{\epsfig{file=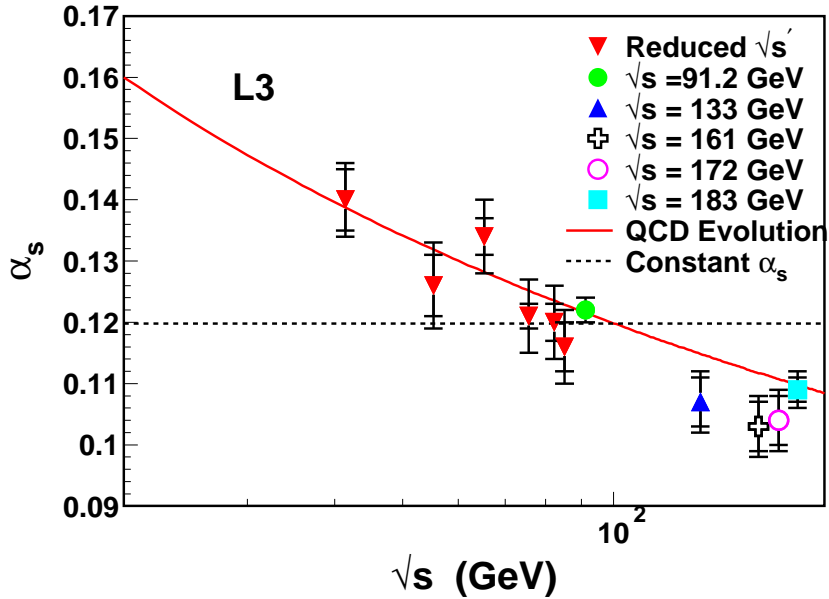, height = 8.5 cm }}
\end{center}
\caption[]{\as measurements from L3 data as a function of $E_{cm}$.
           The solid and dashed lines represent fits with an energy 
           evolution as predicted by QCD and with constant \asx, 
           respectively.}
\label{l3runana}
\end{figure}


Within the L3 analysis\cite{l3running} the running of \as has been
studied including also \ee annihilation data at reduced center of 
mass energies $E_{cm} < M_{z}$. For this purpose, events with hard,
isolated photons have been selected. This high energy photons are
radiated either through initial state radiation or through quark
bremsstrahlung, which takes place before the evolution of the 
hadronic shower. The QCD scale is assumed to be the center of mass
energy of the recoiling hadronic system $\sqrt{s^{'}} $, which is 
related to the photon energy $E_{\gamma}$ by 

\begin{equation}
\label{redenergy}
   \sqrt{s^{'}} 
   = \sqrt{s \left( 1 - \frac{2E_{\gamma}}{\sqrt{s}} \right)}
\end{equation}

\nin
The selection provides \as determinations at reduced $\sqrt{s^{'}}$
from 30 to 86 $\gev$. \as has been determined from a QCD fit in matched
\oass and NLLA precision to the distributions of thrust, heavy jet mass, 
wide and total jet broadening. Hadronization corrections have been
calculated by the means of MC models. Although the matched predictions
turned out to be less reliable\cite{delpap} for the determination of
an absolute value for \asmzx, this should be no problem in terms of
the running of the strong coupling, since the errors are fully 
correlated. Fig. \ref{l3runana} shows the \asmz values measured
as a function of $E_{cm}$ together with a fit to the QCD evolution 
equation. The fit yields a $\chi^{2}$ of 16.9 for 10 degrees of 
freedom, corresponding to a confidence level of 0.076, whereas a
model with constant \as yields a $\chi^{2}$ of 91.4  corresponding 
to a confidence level of $2.9 \times 10^{-13}$.


\section{Status of the Strong Coupling}

Measurements of the strong coupling are available from a large number
of different reactions. Some problems arise, when the individual results
are combined in order to calculate a global average. First, a global 
average of \asmz contains a certain subjective element in the way the
input data are selected. There are for example a large number of
measurements in \ee annihilation at different center of mass energies
which are however expected to be strongly correlated. The fact, that 
the exact correlation pattern between different measurements is unknown 
suggests a pre-clustering of the input data in order to achieve a 
balanced mixture of measurements from different reaction types, which 
are then hopefully less correlated.     
Further problems arise due to the dominance of theoretical 
uncertainties within the \as determinations. Most experiments try
to calculate uncertainties due to missing higher order contributions
by the means of the variation of the renormalization scale, however,
the range within the scale should be varied is quite arbitrary and 
different for each experiment. Therefore the resulting uncertainties 
on \asmz are arbitrary to a large extent as well.

Therefore, three different numbers will be given within the following 
considerations. The first number is a simple unweighted mean, which 
has within this context the advantage to ignore the doubtful scale
ambiguity errors at all, however, different experimental uncertainties 
are ignored as well. The second number will be a simple weighted average,
which does not account for the (unknown) correlations between the
different \as measurements. Also an estimate for a correlated weighted
average as introduced in\cite{coraver} will be given. This method tries
to estimate the covariance matrix by assuming a common correlation 
between the measurements, described by a single parameter $\rho_{\rm eff}$.
The method applies, if $\chi^{2} < n_{df}$. Then, the measurements are 
assumed to be correlated and $\rho_{\rm eff}$ is adjusted until the 
expectation $\chi^{2} = n_{df}$ is satisfied. In general, this method
yields a conservative error estimate. The uncertainty of the average 
value might be quite large, if (e.g. theoretical) uncertainties are 
overestimated for some of the measurements included. Too small errors
for the individual measurements yield in an underestimation of the
correlation between the measurements.


\subsection{Status of \bfas (early 1995)} 


\begin{figure} 
\begin{center}
\mbox{\epsfig{file=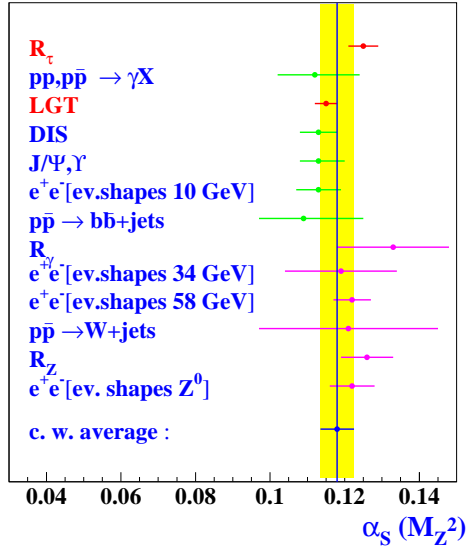, height = 8.5 cm }}
\end{center}
\caption[]{Summary of \asmz measurements at the beginning of 
           1995\cite{schmelling}. The measurements are ordered
           according to the energy of the process. The shaded
           band represents the uncertainty of the correlated
           weighted average.}
\label{status95}
\end{figure}



\begin{table}[b]
\begin{center}
\begin{tabular}{lcc}
\hline 
  value & \asmz & $\chi^{2}/n_{df}$ \\
\hline
  simple mean             & $0.1185 \pm 0.0072 $ &          \\
  simple weighted average & $0.1180 \pm 0.0022 $ & 6.4 / 10 \\ 
  correlated w. average   & $0.1180 \pm 0.0045 $ &          \\ 
\hline

\end{tabular}
\end{center}
\caption{A global average for \asmz measurements representing
         the status in early 1995, estimated using three different 
         methods.}
\label{aver95}
\end{table}


In order to illustrate the enormous progress achieved within 
the last three years, this overview is started with a summary 
of \asmz measurements\cite{schmelling} from 1995. Fig. \ref{status95}
shows a graphical overview of the different measurements.
There were apparently two problems with the measurements shown. 
First the \as measurements from Lattice Gauge Theory (LGT) 
(\asmz $=0.115 \pm 0.003)$ and from hadronic $\tau$-decays 
(\asmz $=0.125 \pm 0.004)$, which claimed both to be the most precise, 
yielded a large difference. They have been ignored in the global average, 
motivated by the fact that the LGT value has been unstable in the past 
and the $\tau$-decay value due to the controversial discussion about 
the validity of some specific theoretical assumptions. The second 
problem was, that the \asmz values were clustered into two different 
groups of low and high energy ($Q \gtrsim 25 \gev$) measurements 
(with the exception of \asmz from $\tau$-decays). The difference 
observed gave some input for speculations about new physics occurring 
at this energy scale. The overall scatter of the individual 
measurements is quite large (see table \ref{aver95}), the uncertainty
calculated from a simple weighted average clearly underestimates the 
true uncertainty, however the correlated weighted average seems to be 
appropriate here.


\subsection{Status of \bfas today} 
 

\begin{figure} 
\begin{center}
\mbox{\epsfig{file=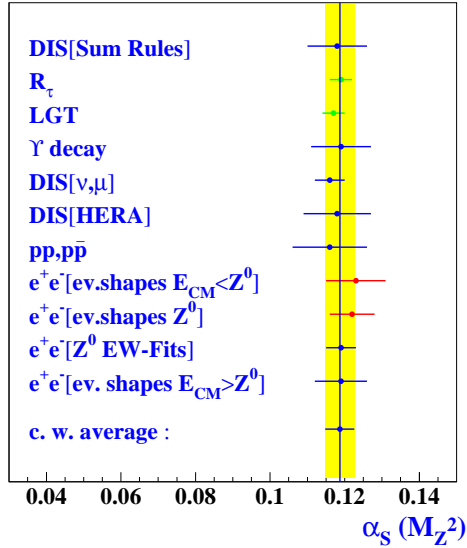, height = 8.5 cm }}
\end{center}
\caption[]{Summary of \asmz measurements. Status after the ICHEP'98
           conference.}
\label{statustoday}
\end{figure}



\begin{table}[b]
\begin{center}
\begin{tabular}{lcc}
\hline 
  value & \asmz & $\chi^{2}/n_{df}$ \\
\hline
  simple mean             & $0.1187 \pm 0.0022 $ &           \\
  simple weighted average & $0.1189 \pm 0.0014 $ & 1.84 / 10 \\ 
  correlated w. average   & $0.1189 \pm 0.0039 $ &           \\ 
\hline

\end{tabular}
\end{center}
\caption{A global average for \asmz measurements representing
         the status after the ICHEP'98 conference. The light
         shaded band represents the uncertainty of the correlated
         weighted average.}
\label{avertoday}
\end{table}


In comparison with 1995 the situation now has drastically changed 
(see Fig. \ref{statustoday} for a graphical overview on current
\as measurements). The difference between \asmz from LGT and from
hadronic $\tau$ decays has been largely reduced. Since the \asmz 
measurements from $\tau$ decays revealed a somewhat larger uncertainty
due to the differences from the models employed, a preliminary average 
of \asmz $= 0.119 \pm 0.003 $ has been considered, corresponding to
an average over the three different models introduced before. The
current PDG average value for \asmz from LGT of 
\asmz $= 0.117 \pm 0.003 $\cite{PDGqcd} is in good agreement with 
the value from $\tau$ decays and there is no longer a reason to 
exclude them from the global average. Furthermore, the two 
clusters of \as values observed earlier completely disappeared. 
The global average for \asmz (see table \ref{avertoday} ) 
is nearly the same than in 1995, however the scatter between 
the individual measurements is largely reduced. 
The $\chi^{2}$ is 1.84 for 10 degrees of freedom, indicating 
that the (theoretical) uncertainties might be overestimated. 
However, the procedure for calculating a correlated weighted average 
interprets the small $\chi^{2}$ entirely in terms of correlations 
between the observables, therefore yielding a quite conservative
error estimate. The largest deviation from the global average comes
from the event shape measurements in \ee annihilation. Here, the results
from matched \oass and NLLA predictions in combination with MC corrections
have been used for the global average, since this is the standard method
nowadays and a large number of measurements have been made. But as we 
have seen within the previous discussion, there are serious arguments,
why this predictions yield less precise results than predictions in \oass
in combination with experimentally optimized scales. It is quite instructive
to study the changes on the global average value, when the results from 
matched predictions are replaced. This will be done in the next section.  


\subsection{looking into the future ...} 
\hspace{0.2 cm}

\begin{figure} [t]
\begin{center}
\mbox{\epsfig{file=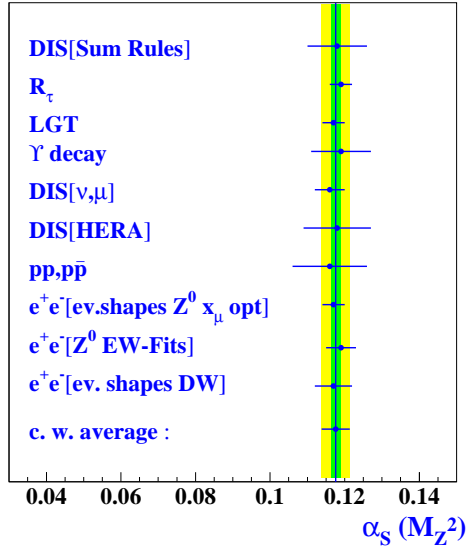, height = 8.5 cm }}
\end{center}
\caption[]{Another summary of \asmz measurements. Here the results
           from matched \oass and NLLA predictions have been replaced
           with results from \oass QCD applying experimentally optimized
           scales, and with results from using DW-model corrections. 
           The light shaded band represents the uncertainty of the 
           correlated weighted average. Additionally the uncertainty 
           of the simple weighted average is indicated by the dark 
           shaded band. }
\label{statusfuture}
\end{figure}



\begin{table} [b]
\begin{center}
\begin{tabular}{lcc}
\hline 
  value & \asmz & $\chi^{2}/n_{df}$ \\
\hline
  simple mean             & $0.1185 \pm 0.0072 $ &          \\
  simple weighted average & $0.1180 \pm 0.0022 $ & 6.4 / 10 \\ 
  correlated w. average   & $0.1180 \pm 0.0045 $ &          \\ 
\hline

\end{tabular}
\end{center}
\caption{A modified global average for \asmz where the results from
         matched predictions have been replaced.}
\label{averfuture}
\end{table}


In a first step, the global results from \ZZ data have been replaced 
by the single result of \asmz $= 0.117 \pm 0.003$ from the DELPHI 
collaboration\cite{delpap}, obtained from \oass predictions in
combination with experimentally optimized scales. Secondly, it has been
assumed, that the DW-model gets established and the results
from \ee data with $E_{cm}\neq M_Z$ have been replaced by 
an average value of current results from DW-model predictions.    
This average has been calculated assuming fully correlated errors which
yields \asmz $= 0.117 \pm 0.005$. (Here, the results of \cite{pedro} have
been ignored, since only statistical errors are given.) See Fig.
\ref{statusfuture} for a graphical view and table \ref{averfuture}
for the average value. The modified result is quite impressive.   
The global average is about 1 \% smaller than before, and the consistency
of the measurements gets further improved. The scatter of the individual
measurements is now only $\pm 0.0012$ and the uncertainty of 
\dasmz $=\pm 0.0038 $ seems now really be too pessimistic. The 
uncertainty obtained from a simple weighted average is about 1 \%
and indicated in Fig. \ref{statusfuture} for illustration reasons.
Clearly, no correlations between the measurements are considered here.
Also the new results introduced here have to be established.
However, if one considers the progress achieved within the last three
years, a 1\% error on \asmz seems to be a realistic 
perspective for the foreseeable future.


\section{Summary}

An enormous progress has been achieved on the determination of \asmz
and its running with the analyses presented at this years summer 
conferences. The importance of adjusting the renormalization scale
has been demonstrated with the analysis of high statistics and
high precision \ZZ data using angular dependent shape observables.  
The observation is confirmed by the analysis of higher moments of
event shapes from LEP high energy data. Comparison of the data with
predictions from matched \oass with NLLA precision revealed an so far
unreported problem, presumably arising due to a mismatch of the
renormalization scales. The DW-model is a novel way in 
understanding non-perturbative power corrections to event shape 
distributions. First results are impressive. Applying the DW-model,
QCD predictions in \oass precision and matched \oass and NLLA 
precision yield similar \asmz values, which are also in good agreement 
with other results from precise \asmz measurements. If hadronization
corrections from MC models are applied instead, a larger deviation of
\asmz is observed for matched predictions, however still compatible 
with the global \asmz average value. The running of \as is clearly
established from LEP \ee data only. All recent precise \asmz 
measurements agree very well with each other, a global average of \\ 

\begin{center}
\asmz $=0.1189 \pm 0.0039$
\end{center}

\nin
has been determined using a correlated weighted average. Replacing
the results from matched \oass and NLLA predictions with the
new results obtained from applying experimentally optimized scales and
from DW model predictions further improves the consistency of the 
global \asmz measurements. A roughly 1 \% error on \asmz seems to be
a realistic perspective for the foreseeable future. 



\end{document}

%% file: commands.tex
\newcommand{\as}{$\alpha_s $ }
\newcommand{\asx}{$\alpha_s$}
\newcommand{\asmz} { $\alpha_s(M_Z^2) $ } 
\newcommand{\asmzx} { $\alpha_s(M_Z^2)$} 
\newcommand{\asmt} { $\alpha_s(m_{\tau}^2) $ } 
\newcommand{\asmtx} { $\alpha_s(m_{\tau}^2)$} 
 
\newcommand{\bfas}{\protect{\boldmath $\alpha_s $ \unboldmath}}

\newcommand{\dasmz}{ $\Delta \alpha_s(M_Z^2) $ }  
\newcommand{\oas}  {$\cal O$($\alpha_s $) } 
\newcommand{\oass} {$\cal O$($\alpha_s^2$) } 
\newcommand{\oasss}{$\cal O$($\alpha_s^3$) } 
\newcommand{\oaSS} {$\cal O$($\alpha_s^4$) } 
\newcommand{\oaSSx} {$\cal O$($\alpha_s^4$)} 
 
\newcommand{\oassx} {$\cal O$($\alpha_s^2$)} 

\newcommand{\MSB}  {$\overline{MS} $ } 
 
\newcommand{\tht} {$\vartheta_T $ } 
\newcommand{\ctt} { $ \cos \vartheta $ } 
\newcommand{\cttgz} { $ \cos \vartheta \ge 0 $ }
\newcommand{\chidf} { $ \chi^{2} / n_{df}$ }

\newcommand{\bftau}{\protect{\boldmath$\tau $ \unboldmath}} 
  
\newcommand{\gev}{\:\mbox{Ge\kern-0.1exV}} 
\newcommand{\mev}{\:\mbox{Me\kern-0.1exV}}

\newcommand{\beq}{\begin{equation}} 
\newcommand{\eeq}{\end{equation}} 
\newcommand{\bea}{\begin{eqnarray}} 
\newcommand{\eea}{\end{eqnarray}}

\newcommand{\ee} {$\mbox{e}^+\mbox{e}^-$ } 
  
\newcommand{\ZZ} { $\rm Z^{0}$ } 
\newcommand{\ZZx} { $\rm Z^{0}$} 
\newcommand{\ZZbf} { $\bf Z^{0}$ }
\def\nin{\noindent}